\documentclass[aps,prl,10pt,twocolumn,superscriptaddress]{revtex4-2}

\usepackage{amsmath,amssymb}
\usepackage{enumitem}
\usepackage{graphicx}
\graphicspath{ {Plots/} }
\usepackage{dcolumn}
\usepackage{bm}
\usepackage{wasysym}

\usepackage{xcolor}
\usepackage{comment}
\makeatletter
\renewcommand\frontmatter@abstractwidth{\dimexpr\textwidth-0in\relax}
\makeatother

\newcommand{\beq}{\begin{equation}}
\newcommand{\eeq}{\end{equation}}
\newcommand{\bea}{\begin{eqnarray}}
\newcommand{\eea}{\end{eqnarray}}

\begin{document}

\title{\LARGE \bf Dirac lines and loop at the Fermi level in the Time-Reversal Symmetry Breaking Superconductor LaNiGa$_2$}

\author{Jackson R. Badger}
	\affiliation{Department of Chemistry, University of California, Davis, California $95616$, USA}
\author{Yundi Quan}
\affiliation{Department of Physics and Astronomy, University of California, Davis, California $95616$, USA}
\affiliation{Present address: Department of Physics, University of Florida, Gainesville, Florida 32611, USA}
\affiliation{Present address: Department of Materials Science and  Engineering, University of Florida, Gainesville, Florida 32611, USA}
\affiliation{Present address: Quantum Theory Project, University of Florida, Gainesville, Florida 32611, USA}
\author{Matthew C. Staab}
	\affiliation{Department of Physics and Astronomy, University of California, Davis, California $95616$, USA}
\author{Shuntaro Sumita}
	\affiliation{Condensed Matter Theory Laboratory, RIKEN CPR, Wako, Saitama 351-0198, Japan}
\author{Antonio Rossi}
	\affiliation{Department of Physics and Astronomy, University of California, Davis, California $95616$, USA}
\affiliation{Present address: Advanced Light Source, Lawrence Berkeley National Laboratory, Berkeley, California 94720, USA}
\author{Kasey P. Devlin}
	\affiliation{Department of Chemistry, University of California, Davis, California $95616$, USA}
\author{Kelly Neubauer}
	\affiliation{Department of Physics and Astronomy, University of California, Davis, California $95616$, USA}
\author{Daniel S. Shulman}
    \affiliation{ Department of Physics, University of California, Berkeley, California, 94720, USA}
\author{James C. Fettinger}
	\affiliation{Department of Chemistry, University of California, Davis, California $95616$, USA}
\author{Peter Klavins}
	\affiliation{Department of Physics and Astronomy, University of California, Davis, California $95616$, USA}
\author{Susan M. Kauzlarich}
	\affiliation{Department of Chemistry, University of California, Davis, California $95616$, USA}
\author{Dai Aoki}
	\affiliation{IMR, Tohoku University, Oarai, Ibaraki $311-1313$, Japan}
\author{Inna M. Vishik}
	\affiliation{Department of Physics and Astronomy, University of California, Davis, California $95616$, USA}
\author{Warren E. Pickett}
	\affiliation{Department of Physics and Astronomy, University of California, Davis, California $95616$, USA}
\author{Valentin Taufour}
	\affiliation{Department of Physics and Astronomy, University of California, Davis, California $95616$, USA}

\begin{abstract}
\textbf{
Unconventional superconductors have Cooper pairs with lower symmetries than in conventional superconductors. In most unconventional superconductors, the additional symmetry breaking occurs in relation to typical ingredients such as strongly correlated Fermi liquid phases, magnetic fluctuations, or strong spin-orbit coupling in noncentrosymmetric structures. In this article, we show that the time-reversal symmetry breaking in the superconductor LaNiGa$_2$ is enabled by its previously unknown topological electronic band structure. Our single crystal diffraction experiments indicate a nonsymmorphic crystal structure, in contrast to the previously reported symmorphic structure. The nonsymmorphic symmetries transform the $k_z=\pi/c$ plane of the Brillouin zone boundary into a node-surface. Band-structure calculations reveal that distinct Fermi surfaces become degenerate on the node-surface and form Dirac lines and a Dirac loop at the Fermi level. Two symmetry related Dirac points remain degenerate under spin-orbit coupling. ARPES measurements confirm the calculations and provide evidence for the Fermi surface degeneracies on the node-surface. These unique topological features enable an unconventional superconducting gap in which time-reversal symmetry can be broken in the absence of other typical ingredients. LaNiGa$_2$ is therefore a topological crystalline superconductor that breaks time-reversal symmetry without any overlapping magnetic ordering or fluctuations. Our findings will enable future discoveries of additional topological superconductors.
}
\end{abstract}


\maketitle

\section{Introduction}
The combination of superconductivity with topology is expected to exhibit new types of quasiparticles such as non-Abelian Majorana zero modes~\cite{Kopnin1991PRB,Read2000PRB}, or fractional charge and spin currents~\cite{Volovik1989JPCM}, and provide new platforms for quantum computation technologies~\cite{Kitaev2003AP}. Topological superconductivity can be artificially engineered in hybrid structures~\cite{Fu2008PRL,Mourik2012Science,Nadj-Perge2014Science,Fatemi2018Science,Sajadi2018Science} or it can exist intrinsically in certain unconventional superconductors~\cite{Wray2010NP,Kobayashi2016PRB,Kidwingira2006Science,Pustogow2019Nature,Zhang2018Science,Ran2019Science}. In most intrinsic topological superconductors, the unconventional nature of superconductivity originates from the proximity to magnetic instabilities or strong electronic correlations~\cite{Ran2019Science}.

We report that the time-reversal symmetry breaking superconductor LaNiGa$_2$ derives its unconventional superconducting pairing from the previously unknown existence of Dirac lines and Dirac loop in the normal state. These features are pinned at the Fermi energy where they impact low energy properties including superconductivity. The rich topology of the electronic structure originates from the nonsymmorphic symmetry that guarantees band degeneracies, which in turn, enable interband and/or complex superconducting order parameters that can break time-reversal symmetry. Our results illustrate a novel method towards realizing intrinsic (single-material) topological superconductivity wherein the underlying space group symmetry intertwines the topology with the unconventional superconductivity.

The centrosymmetric superconductor LaNiGa$_{2}$ was previously known to break time-reversal symmetry when it becomes superconducting below $T_{\textrm{sc}}=2$\,K~\cite{Hillier2012}. Subsequent penetration depth, specific heat, and upper critical field studies showed nodeless multigap behavior~\cite{Weng2016PRL}, in contradiction with possible single-band spin-triplet pairing~\cite{Weng2016PRL,Ghosh2020PRB}. All previous experimental investigations were limited to polycrystalline samples and theoretical considerations were based on the previously reported symmorphic $Cmmm$ (No.~$65$) space group~\cite{Yarmolyuk1982}. We reveal that single crystal X-ray diffraction (SCXRD) analysis improves upon previous powder X-ray diffraction (PXRD) work and properly assigns LaNiGa$_{2}$ to a nonsymmorphic $Cmcm$ (No.~$63$) unit-cell. Difficulty discerning the difference between $Cmmm$ and $Cmcm$ from PXRD data has historical precedent~\cite{Makarov1959SPC,Oikawa1996JPSJ}.

\begin{figure*}[ht!]
	\centering
	\includegraphics[width=\textwidth]{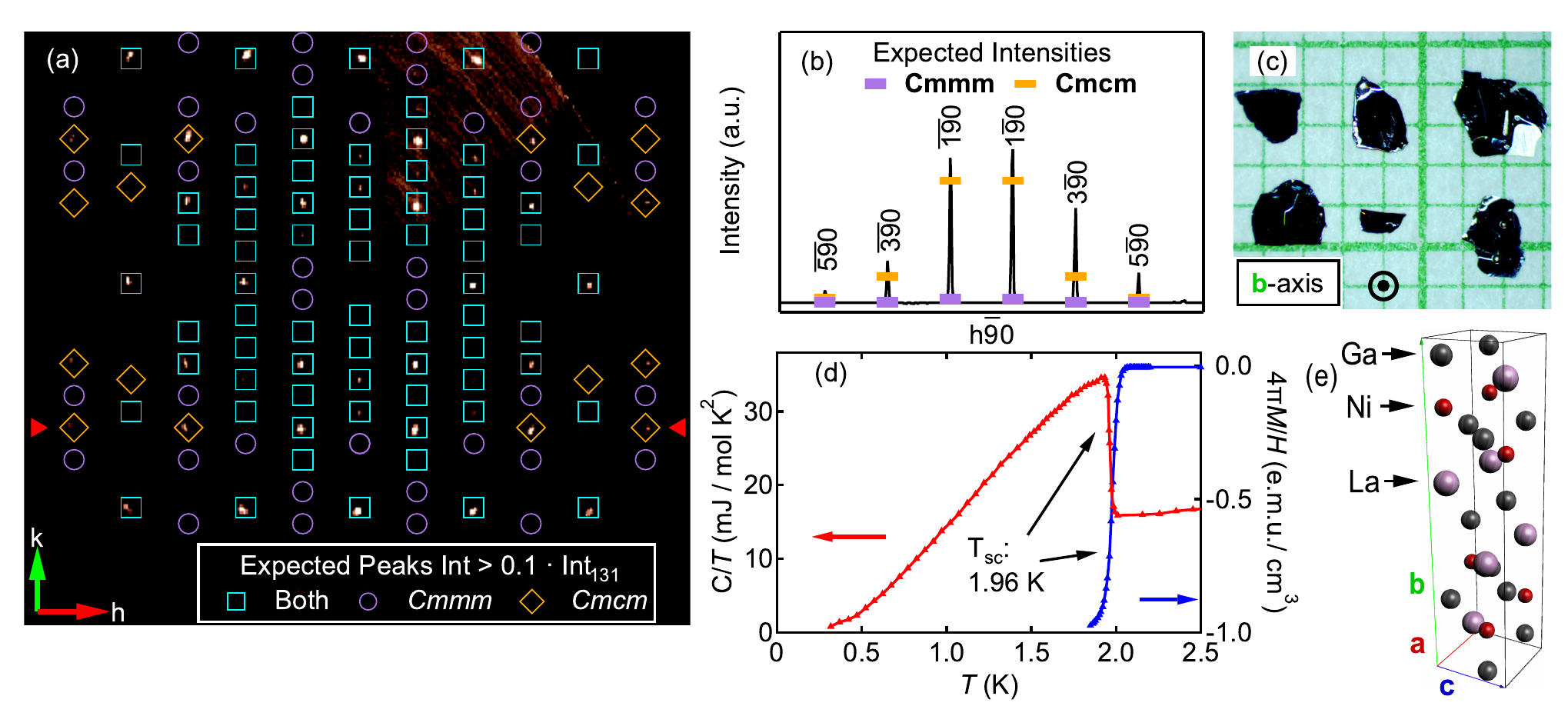}
	\caption{(a) Compiled precession image of the $hk0$ plane from a SCXRD data set of LaNiGa$_{2}$. Overlaid are the predicted diffraction spots with a normalized intensity above $0.1$. The diffraction spots which are expected for both the original $Cmmm$ \cite{Yarmolyuk1982} and our newly proposed $Cmcm$ structures are denoted by the teal squares. While the spots which are only expected for the $Cmmm$ and $Cmcm$ structures are shown by the purple circles and orange diamonds, respectively. (b) A normalized linear cut of the $hk0$ precession image along the $h\overline{9}0$, shown by the red triangles in (a). All intensity data, raw and theoretical, from (a) and (b) are normalized to the $131$ peak. (c) Picture of representative single crystal samples of LaNiGa$_{2}$. The plate-like samples have the $b$-axis normal to the surface of the crystals. (d) A complete superconducting transition is observed with zero-field specific heat capacity ($C/T$) and zero-field cooled magnetic susceptibility ($4\pi\,M/H$) data. Shown by the red and blue curves, respectively. The susceptibility was collected with an external magnetic field of $\mu_0H=1$\,mT. (e) The $Cmcm$ unit cell for LaNiGa$_{2}$ with the same orientation as the BZ in Fig.~\ref{fig:BandDispersion}}
	\label{fig:Precession}
\end{figure*}

The nonsymmorphic symmetries of this new unit cell transform the $k_{z}=\pi/c$ plane of the Brillouin zone (BZ) into a node-surface which hosts four-fold degenerate bands~\cite{Liang2016PRB}. Here, the band degeneracies form two distinct Dirac crossings between two sets of Fermi surfaces (FSs) precisely \textit{at} the Fermi level, independent of chemical potential position. There are fluted lines closed by BZ periodicity and a closed loop. Of special note is that the Dirac loop contains two points which are protected against splitting from spin-orbit coupling (SOC). These ``touchings'' are shown from our band structure calculations, along with angle-resolved photoemission spectroscopy (ARPES) data. 

We note that, among non-magnetic materials and outside of intercalated Bi$_2$Se$_3$, no other time-reversal symmetry breaking superconductor has been shown to exhibit a topological band structure (see Supplementary Table~S$4$). Thus making LaNiGa$_{2}$ unique amongst this small set of bulk superconductors. Lastly, we discuss the impact of the topology of LaNiGa$_{2}$ as a natural platform for interband pairing and/or complex superconducting order parameter that  can  break  time-reversal symmetry.

\section{Results and Discussion} 
\subsection{Structural Characterization}

Single-crystalline samples of LaNiGa$_{2}$ were successfully grown with a Ga deficient self-flux technique. Details about the growth are contained within the materials and methods section below. Highly reflective, plate-like crystals were produced as shown in Fig.~\ref{fig:Precession}(c).

SCXRD data were collected on several samples and each dataset resolved to a LaNiGa$_{2}$ unit cell with a $Cmcm$ space group (see Supplementary Fig.~S$2$ and Tables~S$1$ and S$2$). Given the inherent similarities between the previously reported $Cmmm$ structure~\cite{Yarmolyuk1982} in real space, nearly all diffraction spots within the reciprocal space are predicted by both structures (see Supplementary Fig.~S$1$ for PXRD LeBail fittings using each structure and Fig.~S$3$ for a real space comparison between  the two structures). This is especially true for the most intense, low-angle diffraction spots. There are, however, a few observable differences amongst the weakly-diffracting high-angle spots that are sufficient to differentiate the two structures, as shown from the compiled $hk0$ precession image in Fig.~\ref{fig:Precession}(a). 

\begin{figure*}[t]
	\centering
	\includegraphics[width=\textwidth]{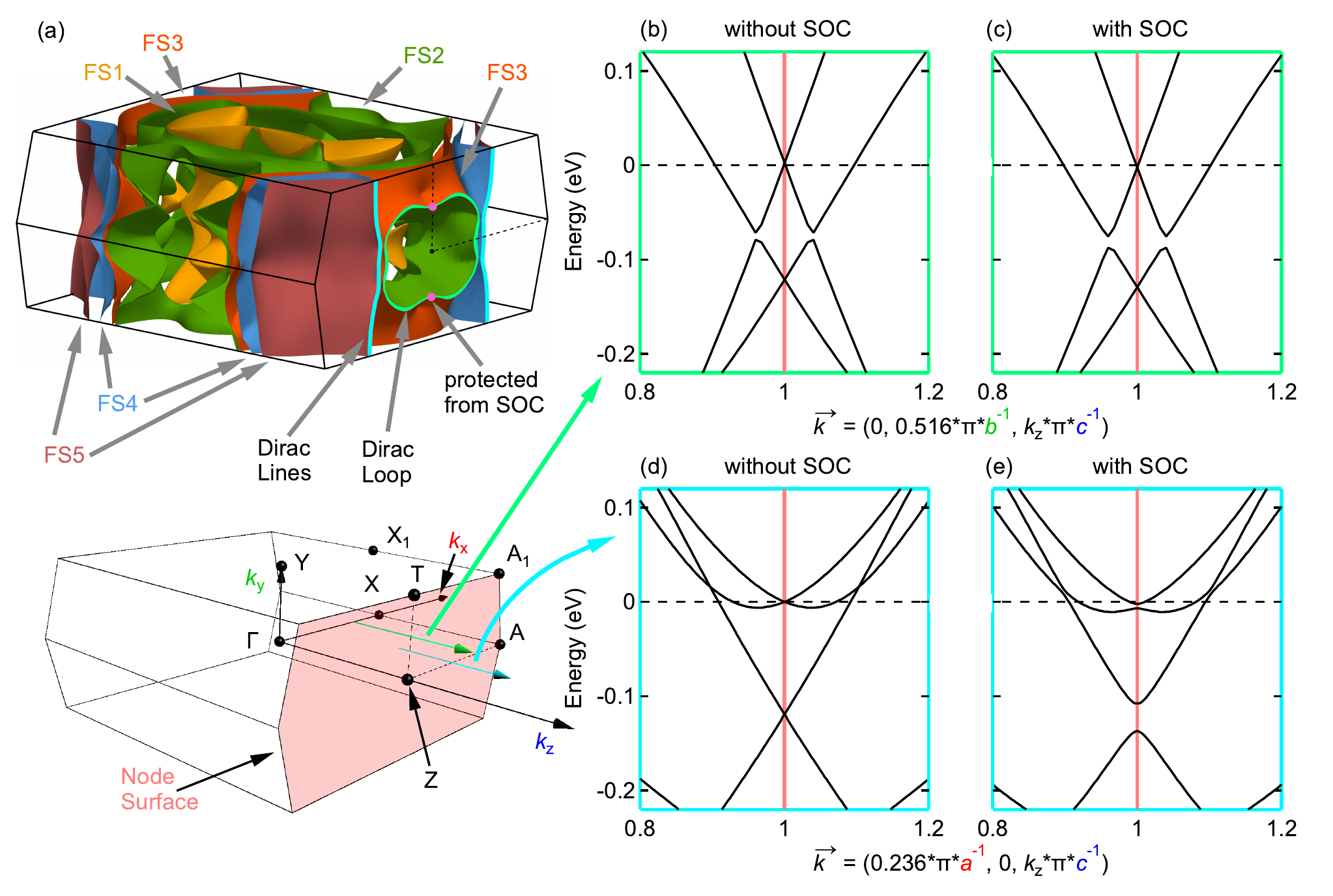}
	\caption{(a) Fermi surfaces within the BZ that highlight the Dirac lines (blue  lines) and Dirac loop (green line) on the node-surface. The crossings along $Z-T$ (magenta dots) are protected from SOC. Below is the BZ showing several high symmetry points and highlights the node-surface (red plane). The green arrow, $k_{y}=0.516\pi/b$, shows where FS$2/3$ become degenerate on the node-surface. The Dirac crossing is shown to remain with (c) and without SOC (b), where the SOC contribution to anticrossing is seen to be very small. The blue arrow shows the dispersion along $k_{x}=0.236\pi/a$ without SOC (d) shows the Dirac lines between FS$4/5$. Once SOC is added (e), the crossing becomes gapped at the node-surface.}
	\label{fig:BandDispersion}
\end{figure*}

These discrepancies are highlighted by the differing expected intensities along the normalized $h\bar{9}0$ line, red arrows in Fig.~\ref{fig:Precession}(a), between the two structures (Fig.~\ref{fig:Precession}(b)). It should be noted that all peaks along this line from both models are displayed, regardless of intensity. The insufficient intensities for the $Cmmm$ model at these high-angle spots reveal that the previously reported structure inadequately matches the observed diffraction data for this material. The data indicates that the more accurate structure for this material is that of the $Cmcm$ space group.

While this new space group remains centrosymmetric, it adds nonsymmorphic symmetries: a $c$ glide plane perpendicular to the $b$-axis and a $2_{1}$ screw axis, $\mathcal{S}^{z}_{2}$, along the $c$-axis. First we will discuss the influence of the new symmetry operations on the electronic structure and later we will examine the ramifications on the superconducting pairing state.

\begin{figure*}[t]
	\centering
	\includegraphics[width=\textwidth]{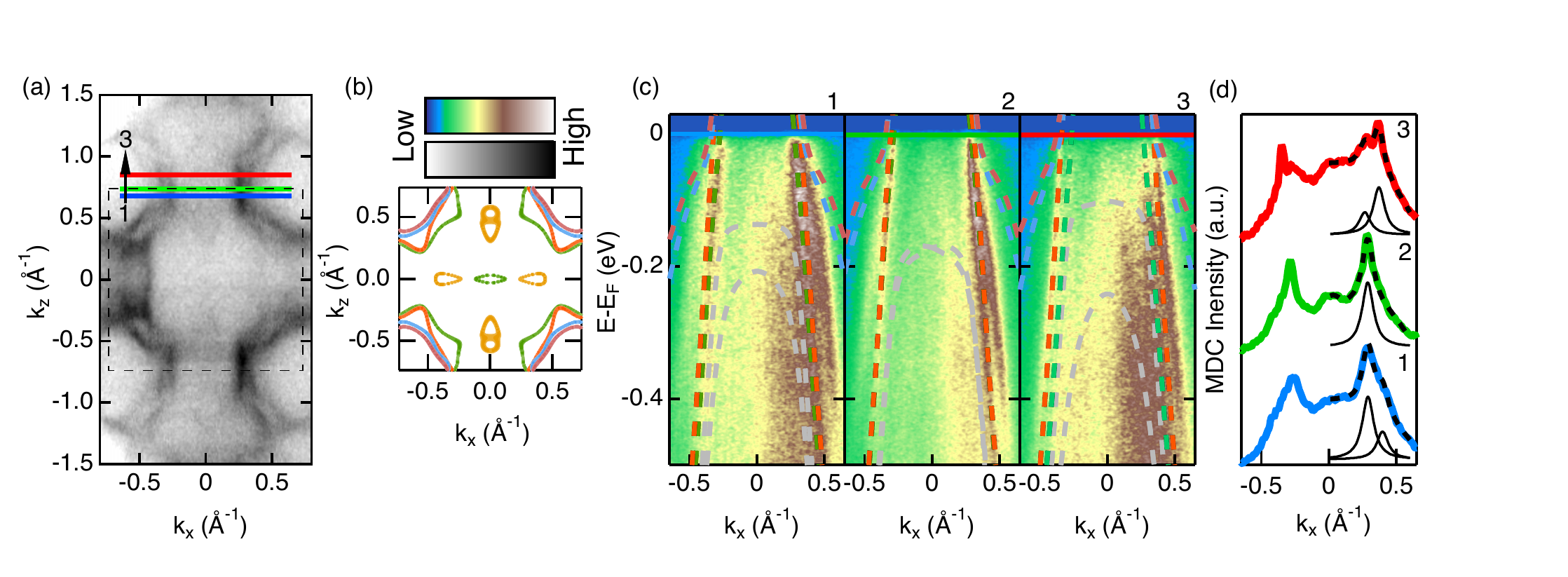}
	\caption{ARPES characterization of LaNiGa$_{2}$. (a) Constant energy map an integration window of $\pm\,10$\,meV around $E_{F}$. The black dotted line indicates the boundary of the BZ in the $k_{y}=0$ plane. The blue, green, and red solid horizontal lines indicate cuts ($1$), ($2$), and ($3$) in panel (c), respectively. (b) Calculated FSs on the $k_y=0$ plane with the colors corresponding to FSs in Fig.~\ref{fig:BandDispersion}(a). See Supplementary Fig.~S$10$(c) for an overlay of the ARPES and the calculated FSs on the $k_y=0$ plane. (c) Energy vs momentum spectra along cuts indicated in (a). The dotted lines are the overlay of DFT calculations and the colors show which FSs are associated with each band. The grey bands are low energy bands which do not cross $E_F$. (d) MDCs at $E_{F}$ from the cuts in panel (c). Spectra are fit to one (two) Lorentzian peaks (dotted black curve) for cuts $2$ ($1$,$3$), with a Gaussian background centered on $k_{x}=0$ (red). The black curves below the experimental data are the individual Lorentzian peaks marking where the bands cross $E_F$. The peak width for the Gaussian fit on $2$ is a free fitting parameter and fixed width for cuts $1$ and $3$.}
	\label{fig:ARPES}
\end{figure*}

\subsection{Electronic Structure and ARPES Data}

Despite the new structure, Fig.~\ref{fig:BandDispersion}(a) shows that there remain $5$ FSs~\cite{Singh2012,Ghosh2020PRB}. Highlighted with the previous space-group, there are several regions within the BZ where pairs of FSs are parallel and quasi-degenerate~\cite{Weng2016PRL,Ghosh2020PRB}. A crucial difference now is that the nonsymmorphic symmetry operations force the previously quasi-degenerate FS pairs to ``touch'' on the $k_{z}=\pi/c$ plane (red plane) in the absence of SOC, turning this plane into a node-surface~\cite{Liang2016PRB} (see the band structures along $T-Z-A$ in Supplementary Fig.~S$11$ to see bands become degenerate in the $Cmcm$ structure). The combination of $\mathcal{S}^{z}_{2}$, parity operation, and time-reversal symmetry force bands on the node-surface to be $4$-fold degenerate (for the derivation and for the differences in the $Cmmm$ and $Cmcm$ band structures see Supplementary Information). This symmetry enforced degeneracy results in two disjoint sets of Dirac crossings directly at the Fermi level. Both are between bulk bands crossing the node-surface: fluted lines across the BZ face between FS$4/5$ and a closed-loop between FS$2/3$ (highlighted lines top panel Fig.~\ref{fig:BandDispersion}(a)). That these crossings occur {\it at} the Fermi level make LaNiGa$_2$ uncommon compared to other superconductors with topologically non-trivial band structures~\cite{Wray2010NP,Zhao2015NC,Sakano2015NC,Neupane2016NC,Guan2016SA,Du2017NC,Zhang2018Science,Jin2019NPJCM,Fang2019AM,Gao2020PRB} (see Supplementary Table~S$4$).

The Dirac crossings can be observed in the linear band dispersion plots without SOC along $\vec{k}=(0,0.516\frac{\pi}{b},k_z)$ (green arrow) for the Dirac loop (Fig.~\ref{fig:BandDispersion}(b)) and $\vec{k}=(0.232\frac{\pi}{a},0,k_z)$ (blue arrow) for the Dirac lines (Fig.~\ref{fig:BandDispersion}(d)). We note that small shifts of the Fermi energy will shift the $k$-space location of the Dirac lines and loop. However, these features will persist at the Fermi level as long as the FSs cross the node-surface. When accounting for SOC, most band crossings become gapped (by a few to $40$\,meV), as pictured in Fig.~\ref{fig:BandDispersion}(e). Remarkably, the Dirac points between FS$2/3$ survive along the $Z-T$ symmetry line under SOC, as seen in Fig.~\ref{fig:BandDispersion}(c), creating two true-Dirac points at the Fermi level. This protected feature results from the presence of the mirror reflection, $\mathcal{M}_{x}$, along the $Z-T$ line, therefore, remaining $4$-fold degenerate even when accounting for SOC (see Supplementary Information for derivation), illustrating that this degeneracy lies precisely at $E_F$, and is robust rather than accidental.

Single crystals of LaNiGa$_2$ do not naturally cleave perpendicular to the crystallographic $c$-axis, making a direct observation of the Dirac dispersion by ARPES measurements challenging. However, with a photon energy of $144$\,eV we can probe the $k_y=0$ plane and confirm the presence of the band touchings (see Supplementary Fig.~S$10$). Fig.~\ref{fig:ARPES}(a) shows the constant energy map centered at $E_F$ and reveals the most prominent features of the spectra: the ruffled cylindrical bands centered on the BZ corners. Given that, near the corner of the BZ, the calculated FSs are very close to each other (see Fig.~\ref{fig:ARPES}(b)), it is difficult to discern which bands are observed in the ARPES measurements from just this plane. Overlaid on Fig.~\ref{fig:ARPES}(c) are the respective DFT band calculations (dashed lines) which reveal that the most prominent bands in the ARPES data originate from the bands associated with FS$2$ and FS$3$. The three parallel horizontal cuts on and near the node-surface show the band dispersion plots at and below $E_{F}$ (Fig.~\ref{fig:ARPES}(c)). The green line, spectrum $2$, represents the cut exactly on the node-surface, while the blue, spectrum $1$, and red, spectrum $3$, lines are parallel cuts in the first and second BZs, respectively.

Each of these linear cuts was integrated within $50$\,meV of $E_F$ to produce momentum distribution curves (MDCs) shown in Fig.~\ref{fig:ARPES}(d). On the node-surface, spectrum $2$ shows a single clear peak representing the degeneracy of FS$2/3$. Off the node-surface, the MDCs for spectra $1$ and $3$ show that FS$2$ and FS$3$ separate and are no longer degenerate. Thus providing direct evidence for the band degeneracy between FS$2/3$ on the node-surface. As mentioned above, we expect SOC to split the FS$2/3$ crossing on the $k_y=0$ plane of the node-surface. We note, however, that the SOC gap cannot be resolved because the peaks have a smaller calculated momentum separation than the fitted experimental widths. This result is further evidence for the minimal impact of SOC on the electronic structure of LaNiGa$_{2}$ in the normal state.

In the normal state and in the absence of SOC, the $Cmcm$ space group makes LaNiGa$_{2}$ a {\it topological nodal line metal}. The line (or loop) is topological~\cite{Allen2018PC}. Nodal lines (lines or loops of degeneracies) in band structures have been found to be rather common~\cite{Herring1937PR,Jin2019PRM}. However, LaNiGa$_2$ is so far unique in having the nodal lines lie {\it precisely at the Fermi level}. However, this confluence of bands will occur in any nonsymmorphic metal with Fermi surfaces crossing the node-surface where bands are guaranteed to be orbitally degenerate.

\subsection{Pairing model and quasiparticles}

Now we examine the consequences of the $Cmcm$ space group assignment for the superconducting state. LaNiGa$_2$ has low symmetry and previous symmetry analysis based on the $D_{2h}$ point group revealed only 4 possible gap functions that break time-reversal symmetry~\cite{Annett1990AP,Hillier2012}. All of them have nodes inside the BZ, which is incompatible with thermodynamic measurements on polycrystals~\cite{Weng2016PRL}, as well as our heat capacity measurements on single crystals which indicate nodeless fully gapped superconductivity (Fig.~\ref{fig:Precession}(d)). The presence of nonsymmorphic symmetries modifies the nodal behavior on the $k_z=\pi/c$ plane with or without SOC (see our classification in Supplementary Information), but does not provide a scenario for the absence of nodes inside the BZ.
The five FSs in Fig.~\ref{fig:BandDispersion} indicate that the full FS is large and pervasive throughout the zone, thus any superconducting gap nodes in a direction ${\hat k}$ would produce a gap node on the FS and thus be detectable in thermodynamic measurements. This observation limits the possible superconducting states to $A_{1g}$ with or without SOC (see Supplementary Information), but these states do not break time-reversal symmetry. The superconducting properties of LaNiGa$_2$ cannot be understood without involving inter-band pairing~\cite{Weng2016PRL,Ghosh2020PRB}. The topological properties of the normal state now provide a natural platform for such unconventional superconductivity.

\begin{figure*}
	\centering
    \includegraphics[width=0.7\textwidth]{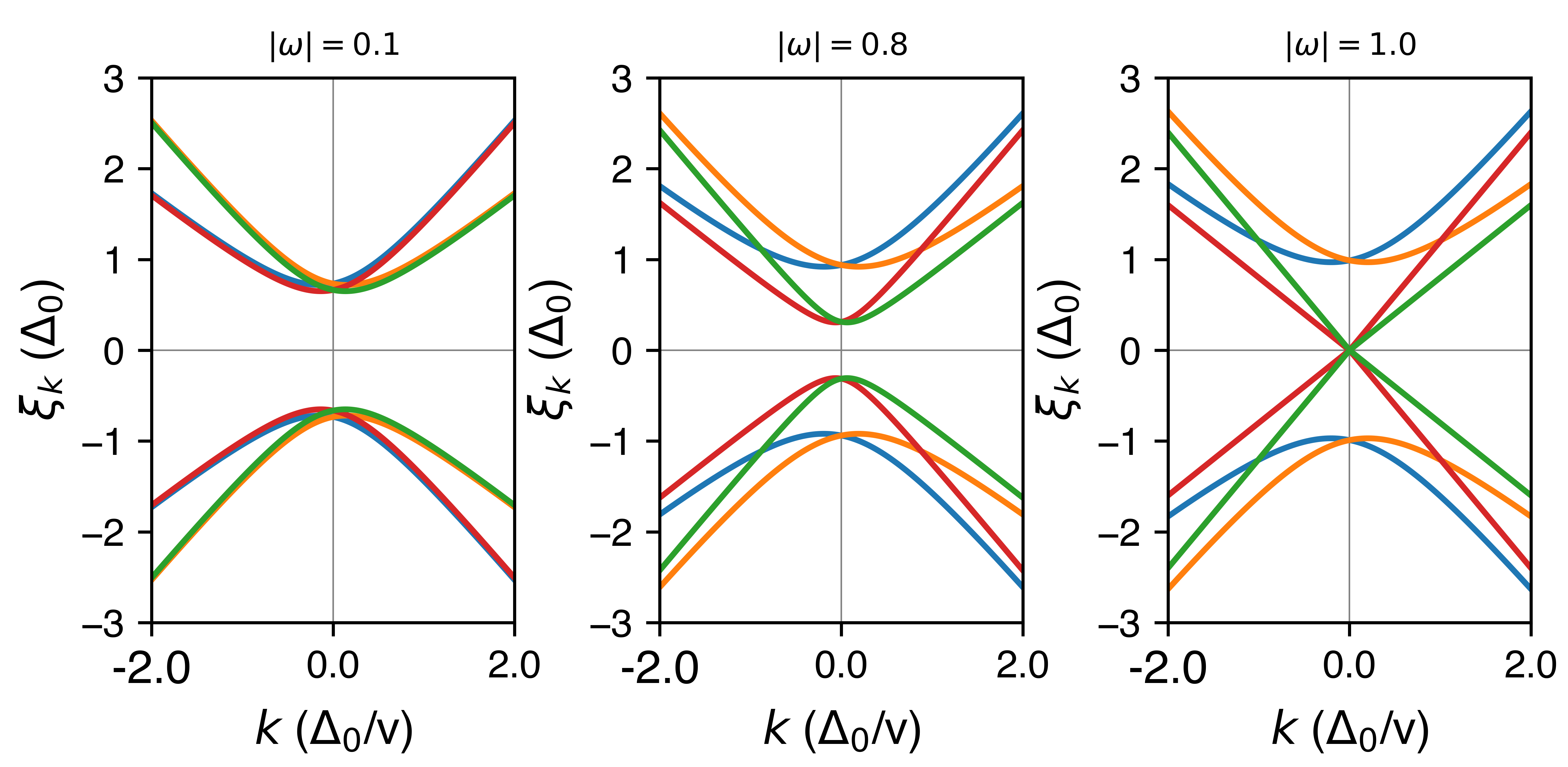}
	\caption{BdG quasiparticle bands near the Dirac point for three values of the $|\vec \omega|=0.1, 0.8, 1.0$. Left to right: gapped, weakly non-unitary to the gapless fully non-unitary limit. Energy units correspond to $\Delta_o=0.7$\,meV, $v=500$ and $\beta=100$, see Eq.\ref{eqn:BdG}.}
	\label{fig:BdG}
\end{figure*}

As mentioned earlier, the Dirac lines and Dirac loop are gapped by SOC, {\it except} for the true-Dirac points surviving on the $Z-T$ line where SOC vanishes. A feature of more interest for the superconducting phase is that, unlike the case for conventional FSs, in Dirac (or Weyl) metals interband transitions persist all the way to zero energy. Any single band model breaks down, and a two-band model is a minimal model~\cite{Yanase2016PRB}. LaNiGa$_2$ thereby becomes an intrinsically two, degenerate and topological, band superconductor. 

If the interband pairs are symmetric in the band index, then the Cooper-pair wave function will have the same symmetry as the intraband pairs do, $s$-wave will be spin-singlet and $p$-wave will be spin-triplet. But if the interband pairs are antisymmetric in the band index, we can have $s$-wave spin-triplet, or $p$-wave spin-singlet pairing while still satisfying the overall fermionic nature required for a superconducting order parameter~\cite{BlackSchaffer2014JPCM}. If both symmetric and antisymmetric pairing exists simultaneously on the node surface (weak SOC) or on the true-Dirac points on the $Z-T$ line (strong SOC), time-reversal symmetry could be broken in two ways: the band (orbital) channel or the spin channel.

In the band-orbital channel, two gap functions (for example s-wave spin singlet $A_{1g}^s$ and s'-wave spin-triplet $A_{1g}^t$) could form a complex combination similar to $s+is'$ to break time-reversal symmetry. Another possibility is to combine several triplet components. For example, the $B_{1g}^t$ triplet will be split by SOC into $A_g$, $B_{2g}$, and $B_{3g}$. The complex combination could also break time-reversal symmetry. However, a first order transition or multiple transitions are expected in these cases~\cite{Weng2016PRL}, but there is no such evidence in our heat-capacity measurements (Fig.~\ref{fig:Precession}(d) and  Supplementary Fig.~S$7$). Time-reversal symmetry breaking in the band-orbital channel is therefore unlikely.

Because of the possibility of $s$-wave spin-triplet pairing on the Dirac lines, loop, or points, time-reversal symmetry could also be broken in the spin-channel from an internally antisymmetric nonunitary triplet pairing (INT) state~\cite{Ghosh2020PRB}.
The power of symmetry analysis is remarkable in the sense that, even with the incorrect space group, the correct point group ($D_{2h}$) already led to the conclusion that the INT pairing is the only reasonable solution~\cite{Weng2016PRL,Ghosh2020PRB}. However, the necessary degeneracy was not identified because of the wrong space group.
An orbital-singlet equal-spin pairing has also been proposed for doped Dirac semimetals~\cite{Kobayashi2015PRL}.
The INT model has also been proposed to explain time-reversal symmetry breaking and fully gapped superconductivity in LaNiC$_2$~\cite{Csire2018EPJB,Ghosh2020PRB,Sundar2021PRB}. While LaNiGa$_2$ and LaNiC$_2$ are compositionally related, our results on LaNiGa$_2$ highlight new significant differences between the two compounds. LaNiC$_2$ has a symmorphic and non-centrosymmetric space group (\textit{Amm2}, No.~$38$), and thus far no topological band-crossing have been reported~\cite{Subedi2009PRB,Laverock2009PRB,Hase2009JPSJ,Yanagisawa2012JPSJ,Wiendlocha2016PRB,Csire2018EPJB,Zhang2018JSNM}. In addition, electrical resistivity measurements under pressure showed the proximity to a different state characterized by a high-energy scale~\cite{Katano2014PRB}, and magnetic penetration-depth measurements under pressure suggested the proximity of a quantum critical point in LaNiC$_2$~\cite{Landaeta2017PRB}. Further studies remain needed on both materials to confirm the validity of the INT model, and the mechanism of time-reversal symmetry breaking. Our discovery of symmetry imposed band crossing, even under SOC, in LaNiGa$_2$, reinforce the relevance of the INT model for this compound, as well as for other nonsymmorphic superconductors.

Breaking of time-reversal symmetry requires a non-unitary triplet pairing potential $\hat{\Delta} = i({\Delta_o{\bf \vec{\eta}}\cdot {\bf \vec{\sigma}}})\sigma_y \otimes i \tau_y$ where the tensor products include the first (spin, $\sigma$ matrices) channel $\sigma=\uparrow,\downarrow$ and the second (band, $\tau$ matrices) channel, with the bands labeled by $m=\pm$ being degenerate along the Dirac loops, taken to be at $k_{\perp}$=0.

Note that the spontaneous vector field $\Delta_o{\vec \eta}$ couples to spin like a magnetic moment. The pairing matrix describes triplet pairing but is antisymmetric in the band channel ($i\tau_y$) to ensure the fermionic antisymmetry of the pair wavefunction~\cite{Weng2016PRL}. The experimentally observed time-reversal symmetry breaking is ensured by the non-unitarity, which is characterized by a nonvanishing real vector ${\vec \omega} = i (\vec{\eta} \times \vec{\eta}^*)$ which satisfies $|{\vec \omega}| \leq |\vec{\eta}|^2 = 1$. A noteworthy difference with previous INT proposals~\cite{Weng2016PRL,Ghosh2020PRB} is that the true two-band situation in LaNiGa$_2$ is enforced by symmetry, and persists in the Bogoliubov - deGennes (BdG) quasiparticle bands. Accounting for the linear band coupling $\beta k_{\perp}$ away from the node-surface, the dispersion of the eight BdG quasiparticles (two bands, two spins, electrons and holes) becomes 
\begin{equation}
\varepsilon_k = \pm \bigl[\sqrt{\{vk_{\perp} - \mu\}^2 + |\Delta_o|^2(1\pm|\vec \omega|)}\pm \beta k_{\perp}\bigr]
\label{eqn:BdG}
\end{equation}
with degenerate eigenvalues on the node-surface of $|\Delta_o|\sqrt{1\pm|\vec \omega|}$. 

The spectrum, shown in Fig.~\ref{fig:BdG}, illustrates the 8-band behavior versus the strength of non-unitarity. The nonsymmorphic operations guarantee that pairs of BdG quasiparticle bands persist in ``sticking together'' on both sides of the gap at $k_{\perp}$=0, thereby retaining topological character. However, now massive points of degeneracy arise in the BdG band structure, unlike the bands of Ghosh {\it et al.}~\cite{Ghosh2020PRB} which retain no degeneracies and hence no topological character. The linear band mixing results in the gap edges lying slightly away from the plane $k_{\perp}=0$. Unit values of $|\vec \omega|$ lead to gaplessness, with unusual Weyl dispersion of the BdG quasiparticles. The measured magnetic moment of $0.012$\,$\mu_B$, if from spin, corresponds to a small conventional band exchange splitting $\Delta_{ex}=m/N(E_F)=1.8$\,meV. This splitting is comparable to (larger than) the superconducting gap $2\Delta_o\sim 3.5-4 k_BT_c \sim 0.7$\,meV, suggesting it may be central to the exotic pairing mechanism.
In contrast to the INT state, conventional $p$-wave spin-triplet superconductivity is expected to lead to high upper critical fields, because equal spin-pairing is not subject to Pauli limiting, and because most proposed $p$-wave superconductors are heavy fermion systems with high orbital limit~\cite{Sheikin2001PRB,Aoki2019JPSJ,Levy2007NP,Aoki2009JPSJUCoGe,Knebel2019JPSJ,Shivaram1986PRL}. LaNiGa$_2$, however, is not a heavy fermion material ($\gamma_{n}=14.1$\,mJ mol$^{-1}$ K$^{-2}$) and interband pairing is suppressed by the application of a strong magnetic field. Thus the upper-critical field in LaNiGa$_2$ remains low (see Supplementary Fig.~S$9$), even though time-reversal symmetry breaking superconductivity was observed at zero field in $\mu$SR experiments.

Our findings reveal that LaNiGa$_2$ is a topological nonsymmorphic crystalline superconductor~\cite{Varjas2015,Wang2016,Shiozaki2016,Yanase2017}. The normal state electronic structure features Dirac lines and Dirac loop at the Fermi energy enforced by nonsymmorphic symmetries, as well as true-Dirac points that retain their degeneracy under SOC. These findings are expected to be common to a large number of materials with similar crystalline symmetries. In general, when topological materials become superconducting, the superconducting state is unconventional. LaNiGa$_2$ was previously reported as a time-reversal symmetry breaking superconductor with evidence for a fully gapped superconducting state, but the topological properties were unknown. The topological character now provides a natural platform for the INT state to exist. Further experiments and theoretical proposals are necessary to further elucidate the time-reversal symmetry breaking mechanism.

Because of the possibility of a fully gapped behavior mimicking conventional superconductivity, many other materials could have been overlooked. Our results on LaNiGa$_2$ motivate the need to characterize the time-reversal symmetry, with zero-field $\mu$SR experiments or Kerr effect, of other crystalline topological metals~\cite{Tang2019,Vergniory2019,Zhang2019} that become superconducting. While LaNiGa$_2$ is the only intrinsic topological material with nodal features at the Fermi level which has been experimentally shown to break time-reversal symmetry in the superconducting state without any overlapping magnetic fluctuations or ordering, other materials could soon be discovered based on our findings.

\section{Materials and Methods}
{\textbf{Sample Preparation.}} Single-crystalline samples of LaNiGa$_{2}$ were grown with a Ga deficient self-flux technique. Ga ($99.99999$\,\%) atomic composition ranged from $32-36\%$ Ga and the remaining percentage was equally split between La ($99.996$\,\%) and Ni ($99.999$\%). Precursor ingots were first synthesized by arc melting all the elements in an argon environment. The ingots were subsequently loaded into an alumina Canfield crucible set~\cite{Canfield2016PM} and sealed in an evacuated quartz ampule. The material was heated up to $1150^{\circ}$C and held at temperature for several hours. The reaction was then slowly cooled down to $800^{\circ}$C over $100$ hours and then quickly centrifuged. Overall, high-quality single crystals were synthesized and characterized (see Supplementary Figs.~S$4-9$ and Table~S$3$).

It was noted that different starting Ga percentages did not produce a noticeable difference in crystal quality, as evaluated by the residual resistivity ratio (RRR). However, larger single crystals (up to $7$\,mm) were obtained in the more Ga deficient syntheses. Additionally, it was also discovered that the superconducting properties were highly sensitive to oxidation throughout the reaction. Lastly, in more Ga deficient growths, below $32\,\%$, no crystals were obtained when the reactions were centrifuged at $800\,^\circ$C.

{\textbf{Crystal Structure Determination.}} Each synthesis was checked to produce the desired phase by PXRD performed on a Rigaku Miniflex with a Cu X-ray source. LeBail fitting was performed using both the $Cmmm$ and $Cmcm$ space groups in GSAS-II~\cite{Toby2013JAC}. Selected samples were aligned using a Laue X-ray diffractometer to distinguish the $a$- and $c$-axes directions. SCXRD data were collected on several samples of LaNiGa$_{2}$ at $100$\,K using a sealed-tube Mo X-ray source on a Bruker Photon $100$ CMOS X-ray diffractometer (Bruker AXS). Across several crystals, obvious twin domains were observed within diffraction space; although not all samples exhibited this. Regardless of the presence of multiple domains, initial unit cell parameters for each sample suggested a $C$-orthorhombic unit cell that matches well with previous reports: $a$\,= $4.2808$\,{\AA}, $b$\,= $17.466$\,{\AA}, and $c$\,= $4.25778$\,{\AA} (ICSD Nos. $634496$ and $634508$)~\cite{Yarmolyuk1982}. The collected frames were integrated using SAINT within APEX3 version $2017.3$. For every crystal that was diffracted, XPREP suggested the centrosymmetric space group $Cmcm$ (No.~$63$) and the structure was refined down to a R value of $0.0288$ using SHELXL-$2018/3$\cite{Sheldrick2015a}. This $Cmcm$ structure is of the BaCuSn$_2$ structure type (ICSD No. $58648$). The precession image was compiled within APEX3. Structure factors for the precession image models were calculated from Visualization for Electronic and Structural Analysis (VESTA) Ver. $3.4.7$~\cite{Momma2011}. 

{\textbf{Physical Property Measurements.}} Low-frequency AC resistivity measurements were measured using a four-probe technique on a Quantum Design Physical Property Measurement System (PPMS) from $300-1.8$\,K. The PPMS was also used to obtain heat capacity data for selected samples using a relaxation technique down to $1.8$\,K. A $^3$He insert for the PPMS allowed for measurements of AC resistivity and heat capacity down to $0.4$\,K. Magnetization measurements were collected in a Quantum Design DC Magnetic Property Measurement System down to $1.85$\,K.

{\textbf{Electronic Structure Methods.}}
Density functional based electronic structures were produced by the precise linearized augmented planewave code {\sc{Wien2k}}~\cite{Blaha2019,Blaha2020JCP} using the generalized gradient functional for exchange and correlation. The sphere sizes were, in bohr: La, 2.50; Ni, 2.40; Ga, 2.12. The plane wave cutoff $K_{max}$ was determined by $RK_{max}=7$, and the $k$-point mesh for self-consistency was $14\times 14\times 14$. Exchange and correlation contributions to the energy and potential were included by using the generalized gradient approximation functional~\cite{Perdew1996PRL}. Effects of spin-orbit coupling were included by using second variation method as implemented in WIEN2k.

{\textbf{ARPES Measurements.}} ARPES measurements were performed at Stanford Synchrotron Radiation Lightsource National Laboratory beamline $5-2$ using a Scientia DA$30$ electron spectrometer. Samples were cleaved \textit{in-situ} at $20$\,K and with a pressure better than $5\times10^{-11}$\,Torr.

{\textbf{Data Availability}} The data that support the findings of this study are available from the authors upon reasonable request.

{\textbf{Acknowledgments}} We thank Rahim Ullah, Li Si, Jianxin Zhu, Junren Shi, Shingo Yonezawa, Makariy Tanatar, Ruslan Prozorov, and Christopher Perez for helpful discussions. The synthesis and characterizations were supported by the UC Laboratory Fees Research Program (LFR-20-653926). V.~T. also acknowledge funding from GIMRT (19F0502). The ARPES work in this manuscript was supported by AFOSR Grant No. FA9550-18-1-0156. Use of the Stanford Synchrotron Radiation Lightsource, SLAC National Accelerator Laboratory, is supported by the U.S. Department of Energy, Office of Science, Office of Basic Energy Sciences under Contract No. DE-AC02-76SF00515. The work of S.~S. is supported by JST CREST Grant No. JPMJCR19T2. This work used the Extreme Science and Engineering Discovery Environment (XSEDE), which is supported by National Science Foundation grant number ACI-1548562. W.E.P and Y.Q. acknowledge support from U.S. National Science Foundation Grant DMR 1607139. K. N. and D. S. S. were supported by the NSF-REU programs PHY-1560482 and PHY-1852581.

\bibliographystyle{apsrev4-2}
\bibliography{Library,biblio}
\end{document}


\title{Supplementary Information: Dirac lines and loop at the Fermi level in the Time-Reversal Symmetry Breaking Superconductor LaNiGa$_2$}

\author{Jackson R. Badger}
	\affiliation{
Department of Chemistry, University of California, Davis, California $95616$, USA}
\author{Yundi Quan}
\affiliation{Department of Physics and Astronomy, University of California, Davis, California $95616$, USA}
\affiliation{Present address: Department of Physics, University of Florida, Gainesville, Florida 32611, USA}
\affiliation{Present address: Department of Materials Science and  Engineering, University of Florida, Gainesville, Florida 32611, USA}
\affiliation{Present address: Quantum Theory Project, University of Florida, Gainesville, Florida 32611, USA}
\author{Matthew C. Staab}
	\affiliation{Department of Physics and Astronomy, University of California, Davis, California $95616$, USA}
\author{Shuntaro Sumita}
	\affiliation{Condensed Matter Theory Laboratory, RIKEN CPR, Wako, Saitama 351-0198, Japan}
\author{Antonio Rossi}
	\affiliation{Department of Physics and Astronomy, University of California, Davis, California $95616$, USA}
	\affiliation{Present address: Advanced Light Source, Lawrence Berkeley National Laboratory, Berkeley, California 94720, USA}
\author{Kasey P. Devlin}
	\affiliation{
Department of Chemistry, University of California, Davis, California $95616$, USA}
\author{Kelly Neubauer}
	\affiliation{Department of Physics and Astronomy, University of California, Davis, California $95616$, USA}
\author{Daniel S. Shulman}
    \affiliation{ Department of Physics, University of California, Berkeley, California, 94720, USA}	
\author{James C. Fettinger}
	\affiliation{
Department of Chemistry, University of California, Davis, California $95616$, USA}
\author{Peter Klavins}
	\affiliation{Department of Physics and Astronomy, University of California, Davis, California $95616$, USA}
\author{Susan M. Kauzlarich}
	\affiliation{
Department of Chemistry, University of California, Davis, California $95616$, USA}
\author{Dai Aoki}
	\affiliation{
IMR, Tohoku University, Oarai, Ibaraki $311-1313$, Japan}
\author{Inna M. Vishik}
	\affiliation{Department of Physics and Astronomy, University of California, Davis, California $95616$, USA}
\author{Warren E. Pickett}
	\affiliation{Department of Physics and Astronomy, University of California, Davis, California $95616$, USA}
\author{Valentin Taufour}
	\affiliation{Department of Physics and Astronomy, University of California, Davis, California $95616$, USA}


\maketitle

\beginsupplement

\section{Powder X-ray Diffraction} 

\begin{figure}[!htb]
	\centering
	\includegraphics[width=\textwidth]{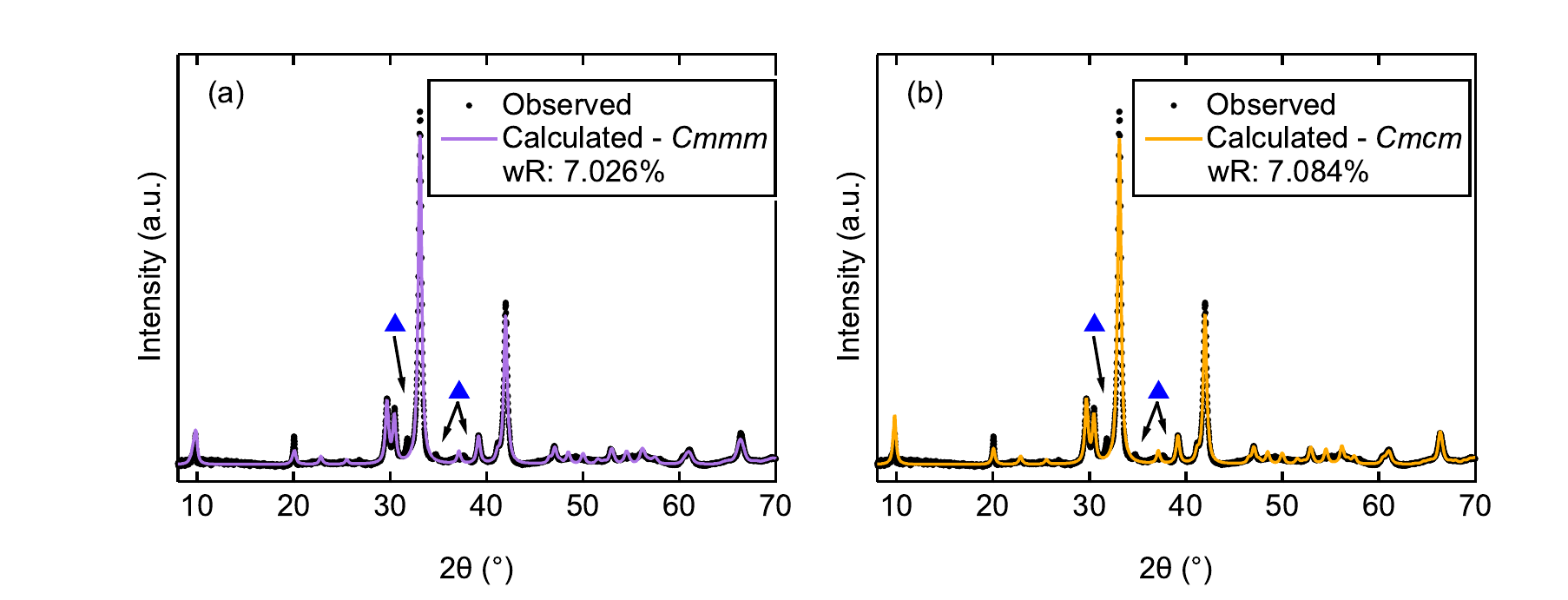}
	\caption{Background subtracted PXRD pattern of LaNiGa$_2$ that was collected from ground single crystals (black dots). The overlaid lines are the calculated models from GSAS-II using the (a) previously reported $Cmmm$ and the (b) new $Cmcm$ structures. The blue triangles denote the peaks from the unidentifiable flux.\label{fig:XRD}}
\end{figure}

Fig.~\ref{fig:XRD} shows the powder X-ray diffraction (PXRD) results with  a LeBail refinement using the previously reported structure \cite{Yarmolyuk1982, Romaka1983} and the new $Cmcm$ structure. The fittings were made using GSAS-II\cite{Toby2013}. In addition to LaNiGa$_{2}$, there was a small amount of impurity flux that is either the LaNiGa or LaNi$_{3}$Ga$_{2}$ phase (blue triangles Fig.~\ref{fig:XRD}). The refined unit-cell parameters from the $Cmmm$ structure are $a=4.278$\,\AA, $b=17.436$\,\AA, and $c=4.271$\,\AA. The refined unit-cell parameters from the $Cmcm$ structure are $a=4.273$\,\AA, $b=17.412$\,\AA, and $c=4.268$\,\AA. Since both the $Cmmm$ and $Cmcm$ structures model well onto the PXRD data (wR$=7.026\%$ and $7.084\%$, respectively), SCXRD is the best technique to experimentally distinguish the correct structure.

\section{Single Crystal X-ray Diffraction and Unit Cell}

From the SCXRD data sets, we conclude that the best structural fit to the data within the capabilities of our instrument is with a $Cmcm$ (No. $63$) space group. This new structure varies in important ways from the previous report, wherein a $Cmmm$ (No. $65$) was suggested. This new space group remains centrosymmetric, with the only symmetry operation difference being the additions of a $c$ glide plane perpendicular to the $b$-axis and a $2_{1}$ screw axis, $\mathcal{S}^{z}_{2}$, along the $c$-axis.

\begin{figure}[b]
	\centering
	\includegraphics[width=\textwidth]{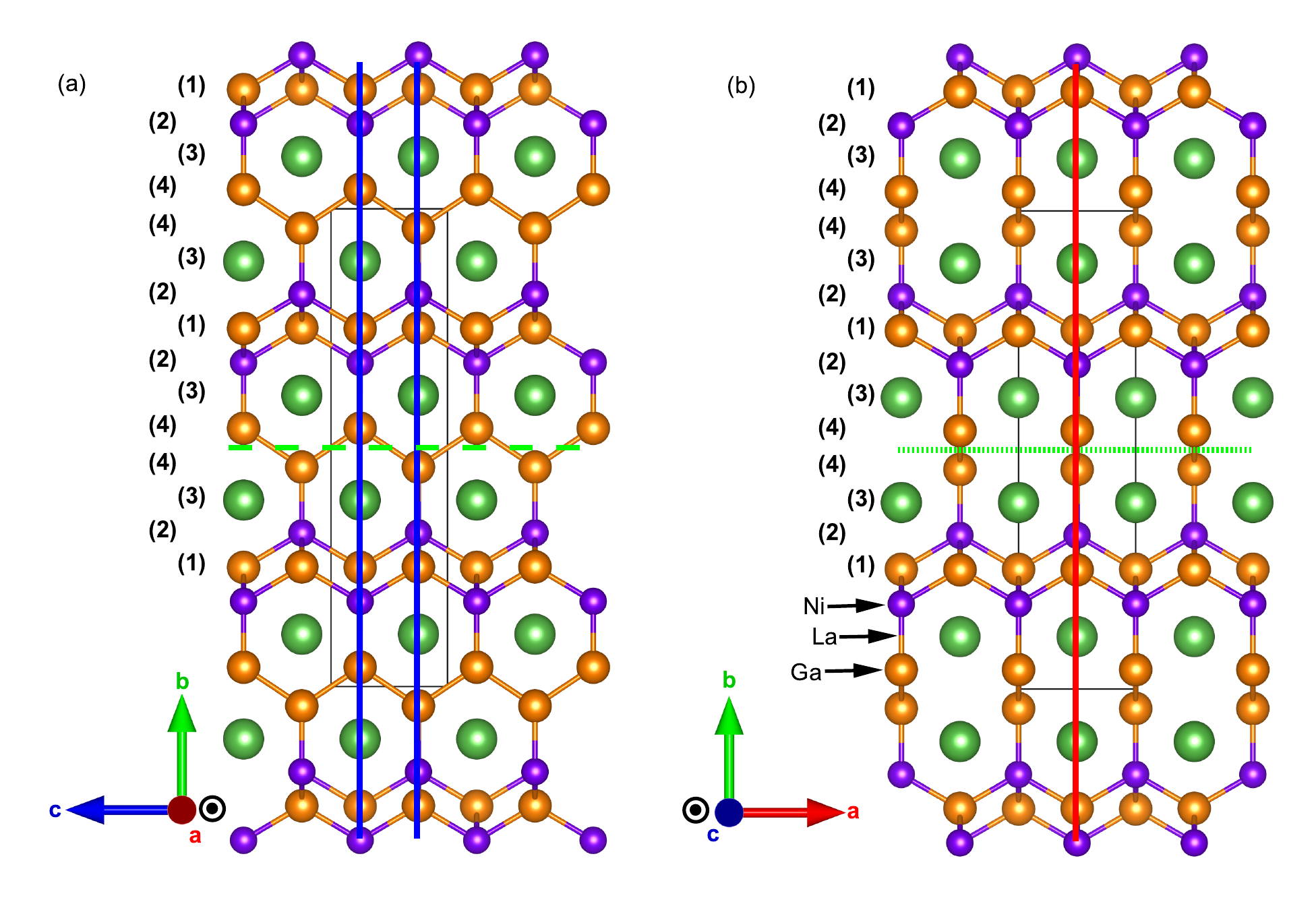}
	\caption{Projection of the new LaNiGa$_2$ structure along the a) $a$- and b) $c$-axis. The numbers denote the layers within the structure: ($1$) body-centered Ga plane, ($2$) Ni, ($3$) La, and ($4$) Ga. The vertical red and blue lines represent the location of the mirror planes, while the dashed green lines denote the $c$ glide plane perpendicular to the $b$-axis.}
	\label{fig:projs}	
\end{figure}

The new crystal structure retains a Z value of four (with two f.u. in the primitive cell) and contains four unique atom positions comprised of one La, one Ni, and two Ga. Details of the SCXRD experiment are highlighted in Table~\ref{tab:Crystal} and atomic positions in Table~\ref{tab:Positions}. When observing the crystal structure projected down the $a$- and $c$-axes (Fig.~\ref{fig:projs}), the structure can be viewed as layers of each element stacking along the $b$-axis. These layers can be described as centrosymmetrically sandwiched together with ($1$) body-centered planes of Ga atoms encasing the motif. Moving inward there are planes of ($2$) Ni atoms, ($3$) La atoms, and ($4$) Ga atoms; each of which are transitionally offset from their respective counterpart plane by ($0$, $0$, $\frac{1}{2}$). These structural projections also show the symmetry elements associated with the $Cmcm$ space group, as highlighted by the colored lines. These operations include the reflection and translation of the $c$ glide plane.

This structure as a whole is bound together by interlayer bonding between the ($1$) Ga - ($2$) Ni, ($2$) Ni - ($4$) Ga, and the inner ($4$) Ga planes. The Ga-Ni bonds allow for the ($1$) body-centered Ga planes to form tetrahedral sheets with the ($2$) Ni atoms as the end caps. This motif is the same as in $\beta$-FeSe layers, except with the $3$d and $4$p elements swapped between the two structures. Differing from $\beta$-FeSe, the capped ($2$) Ni atoms bond to the inner ($4$) Ga layers. The ($4$) Ga is bonded to its offset counterpart to form a Ga-Ga zigzag chain extending in the $c$ direction. Additionally, all these bonds between two ($1$) planes come together to form hexagonal sheets, which are shifted by ($\frac{1}{2}$, $0$, $0$) every-other sheet along the $b$ direction. In whole, these bonds allow for the formation of La channels both between the stacked hexagonal sheets along the $c$ direction and within each hexagon along the $a$ direction.   

\begin{figure}[t]
	\centering
	\includegraphics[width=\textwidth]{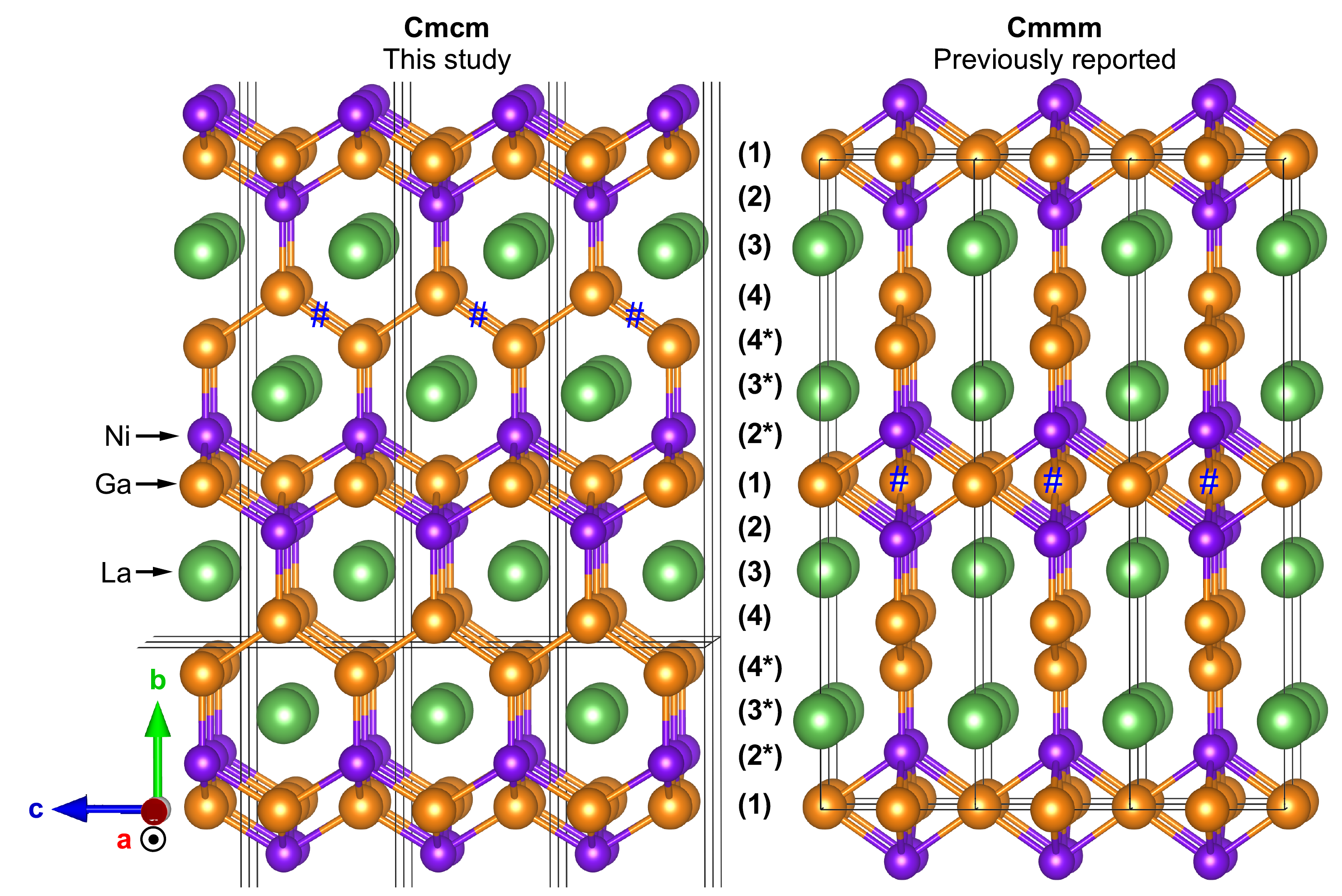}
	\caption{Comparison of the normalized \textit{slabs} between the new $Cmcm$ (left) and the old $Cmmm$ (right) structures. The black lines represent the border of every present unit cell. The inversion center within each unit cell is denoted by the blue $\#$ symbol. The referenced axes apply to both structures.}
	\label{fig:slabs}
\end{figure}

One feature of note is that within the ($2$), ($3$), and ($4$) layers the intraplanar atoms combine to form planar rectangular primitive cells. Despite the different elements, each of these planar cells have separations that are equivalent to the $a$- and $c$-axes. In addition to highlighting the aforementioned layered motif, these structural projections also show the symmetry elements associated with the $Cmcm$ space group, the colored lines. These operations include the reflection and translation of the $c$ glide plane perpendicular to the $b$-axis. 

Not surprisingly, there are many structural features that overlap between the $Cmcm$ and $Cmmm$ unit cells. Beyond the obvious similarities in unit cell dimensions and point group, both structures comprise of the previously mentioned layers and \textit{slabs}. The planes within these repeating motifs appear in the same sequence and the interlayer distances are very similar. When normalized, the largest difference is $0.2$\,\AA. Since both structures contain these \textit{slabs} and every other ($1$) body-centered Ga plane is positionally identical within the $a$-$c$ plane, we can easily compare the contents of the two structures. When these layers are normalized to the same positions, as can be observed in Fig.~\ref{fig:slabs}, we see that within the $Cmcm$ \textit{slab} every other section of the ($2$), ($3$), and ($4$) is shifted by ($\frac{1}{2}$, $0$, $\frac{1}{2}$), denoted by the starred numbers. While the $Cmmm$ \textit{slab} does not exhibit any shifting. This simple translation is the only structural difference and is sufficient to cause the border of the unit cells to shift and subsequently the center of inversion to shift from the central Ga atom within ($1$) plane in $Cmmm$ to half-way between the ($4$) layers in $Cmcm$. With regards to the bonding, these shifted atoms transform the Ga-Ni tetrahedral sheets into pseudo-square planar atom sites and eliminate the translation shift between the hexagonal sheets.

Additionally of note, the $Cmmm$ to $Cmcm$ structures contain the same number of Ni and La sites, but a different number of Ga sites. Transforming from the $Cmmm$ to $Cmcm$, the first and second Ga site locations, comprising the ($1$) plane, converge to a single site location. Although there is an additional site location in $Cmmm$, both the first and the second sites fall on a Wyckoff position with a multiplicity of $2$. In contrast, the converged site in $Cmcm$ falls on a Wyckoff position with a multiplicity of $4$, thus retaining the the stoichiometry between the two structures. 

When thinking about the structural identification saga of LaNiGa$_2$, we could not help but be fascinated by the similarities with that of the superconducting ferromagnet UGe$_{2}$ \cite{Oikawa1996,Boulet1997}. The initial structural misidentification of the two materials follows in nearly the same path, except in opposite directions. Originally thought to have a $Cmcm$ space group, it was not until single crystal structural experiments were performed on UGe$_2$ that the true $Cmmm$ space group was properly identified \cite{Makarov1959,Oikawa1996,Boulet1997}. Beyond similar difficulties with identifying the proper space group, the structural framework of UGe$_2$ and LaNiGa$_2$ (ignoring Ni) were originally identical when they were both $Cmmm$. However, this new space group identification changes the two structures in the same manner as previously mentioned.  

\begin{table*}
	\centering
		\begin{tabular} { l c }
		\hline
			Identification code & JB$10$M$4$FMI       (JF$3040$) \\
			Empirical formula & Ga$2$ La Ni \\
			Formula weight & $337.04$\,g mol$^{-1}$ \\
			Temperature & $100$($2$)\,K \\
			Wavelength & $0.71073$\,\AA \\
			Crystal system & Orthorhombic \\
			Space group & Cmcm \\
			Unit cell dimensions & $a$ = $4.2818(14$)\,\AA \\
				& $b$ = $17.468(6)$\,\AA \\
				& $c$ = $4.2582(15)$\,\AA \\
			Volume	& $318.48(19)$\,\AA$^3$ \\
			Z &	$4$ \\
			Density (calculated) & $7.030$\,Mg/m$^3$ \\
			Absorption coefficient & $35.380$\,mm$^{-1}$ \\
			F($000$) & $588$ \\
			Crystal size & 	$0.159$ x $0.112$ x $0.081$ mm$^3$ \\
			Crystal color and habit &	Silver Block \\
			Diffractometer & Bruker Photon$100$ CMOS \\
			Theta range for data collection & $2.332$ to $27.464^\circ$ \\
			Index ranges & $-5\leq$h$\leq5$, $-21\leq$k$\leq22$, $-5\leq$l$\leq5$ \\
			Reflections collected & $1017$ \\
			Independent reflections	& $234$ [R(int) = $0.0216$] \\
			Observed reflections (I $>2$sigma(I)) & $232$ \\
			Completeness to theta = $25.242^\circ$ &	$100\%$ \\
			Absorption correction & Semi-empirical from equivalents \\
			Max. and min. transmission & $0.0326$ and $0.0072$ \\
			Solution method & SHELXT (Sheldrick, 2014) \\
			Refinement method & SHELXL-$2017/1$ (Sheldric,2017) \\
			 & Full-matrix least-squares on F$^2$ \\
			Data / restraints / parameters & $234$ / $0$ / $18$ \\
			Goodness-of-fit on F$^2$ & $1.345$ \\
			Final $R$ indices [I $>2$sigma(I)] & $R_1$ = $0.0222$, w$R_2$ = $0.0620$ \\
			$R$ indices (all data) & $R_1$ = $0.0223$, w$R_2$ = $0.0621$ \\
			Extinction coefficient & $0.0025(4)$ \\
			Largest diff. peak and hole & $2.287$ and $-1.393$ e. \AA$^{-3}$ \\
		\hline			
		\end{tabular}
	\caption{Crystal data and structure refinement for LaNiGa$_2$.}
	\label{tab:Crystal}
\end{table*}

\begin{table}
	\centering
		\begin{tabular} { c c c c c c }
		\hline
		Atom & Wyckoff Sites & x & y & z & U(eq) \\
		\hline
		La($1$)	& $4$c & $1$ & $0.3903(1)$ & $0.75$	& $15(1)$ \\
		Ga($1$)	& $4$c & $0.5$ & $0.2495(1)$ & $-0.25$ & $19(1)$ \\
		Ga($2$) & $4$c & $0.5$ & $0.4593(1)$ & $0.25$ & $16(1)$ \\
		Ni($1$) & $4$c & $0.5$ & $0.3216(1)$ & $0.25$ & $16(1)$ \\
		\hline			
		\end{tabular}
	\caption{Atomic coordinates and equivalent isotropic displacement parameters (\AA$^{2}x 10^3$) for LaNiGa$_2$. U(eq) is defined as one third of the trace of the orthogonalized U$^{ij}$ tensor.}
	\label{tab:Positions}
\end{table}

\section{Magnetic Susceptibility} 

\begin{figure}[!htb]
	\centering
	\includegraphics{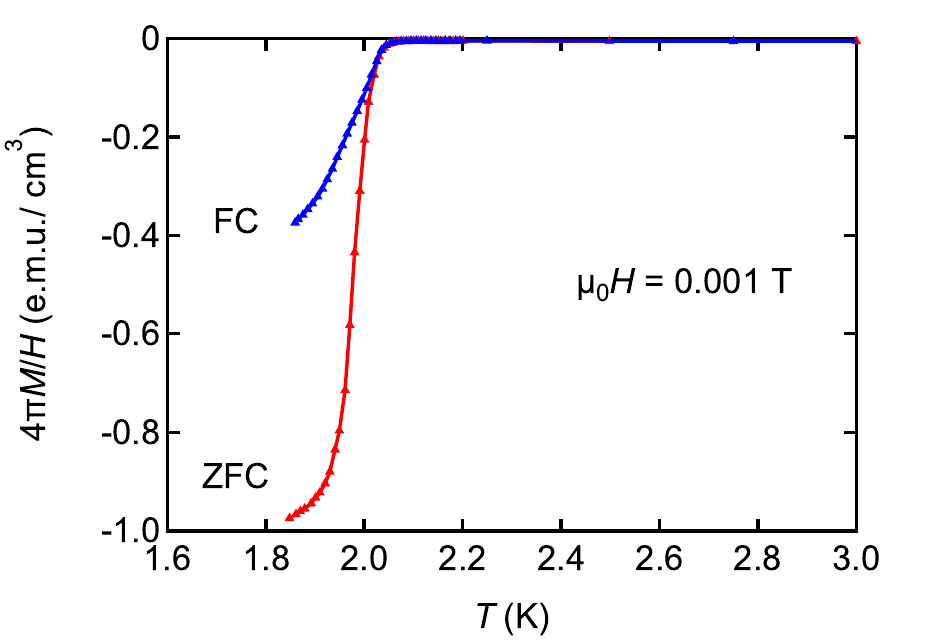}
	\caption{The temperature dependence of the zero-field cooled (ZFC) and field-cooled (FC) magnetic susceptibility ($4\pi\,M/H$) curves for LaNiGa$_2$ with a field of $1$\,mT.}
	\label{fig:Susc}
\end{figure}

Magnetic susceptibility with a magnetic field of $1$\,mT along the $b$-axis is shown in Fig.~\ref{fig:Susc}. Temperature dependence was collected under zero-field cooled (ZFC) and field cooled (FC) conditions. A clear diamagnetic response is observed, with an almost complete expulsion of the external magnetic field for the ZFC curve. The superconducting transition, $T_{sc}^M$, is selected when the material reaches a $90\%$ shielding fraction, at $1.92$\,K. The combination of a sharp transition, $\Delta\,T\,=0.1$\,K, and the magnitude of the diamagnetic response is consistent with the bulk superconductivity confirmed from heat capacity measurements. The transition temperature is in good agreement with what had been previously reported from both AC and DC susceptibility measurements \cite{Zeng2002,Weng2016}. The separation of the ZFC and FC curves indicates a moderate presence of flux pinning in a type-II superconductor, and the scale of difference is less than previous polycrystalline measurements, which is expected for high-quality single-crystal susceptibility measurements with reduced pinning centers.

\section{Electrical resistivity} 

\begin{figure}[!htb]
	\centering
	\includegraphics{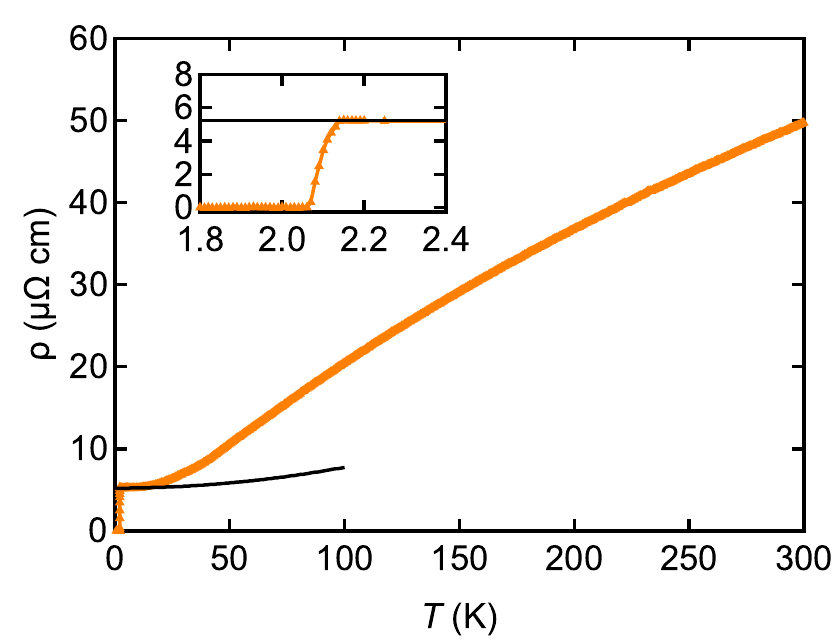}
	\caption{Electrical resistivity ($\rho(T)$) of a representative sample of LaNiGa$_2$. The solid black curve is a fit from the Fermi liquid behaviour of the normal state between $3$-$10$\,K. The inset shows the superconducting transition.}
	\label{fig:Res}
\end{figure}

Electrical resistivity measured in zero-field for a single crystal is shown in Fig.~\ref{fig:Res}. The complete superconducting transition is observed with a $T_{sc}^\rho=2.06$\,K in the inset, while no other anomalies are observed. Fitting the normal state low-temperature region ($3$-$10$\,K) by a Fermi-liquid behavior: \(\rho(T) = \rho_{0} + AT^{2}\) leads to $\rho_0=5.20$\,$\mu\Omega$\,cm and $A=2.54\times10^{-4}$\,$\mu\Omega$\,cm\,K$^{-2}$. The residual resistivity ratio (RRR) for this sample is $9.57$. Both $\rho_0$ and the RRR indicate a higher sample quality than data on polycrystalline samples~\cite{Zeng2002}. Although these values from our samples are slightly worse than a month-long annealed polycrystalline sample~\cite{Weng2016}. Additionally of note, there is a slight negative curvature in the high-temperature region which has been observed in other La-Ni compounds \cite{Goetsch2012, Nakamura2017}, and can arise from $s$-$d$ interband scattering \cite{Mott1964} and electron-phonon coupling.

\begin{figure}
		\centering
		\includegraphics[width=\textwidth]{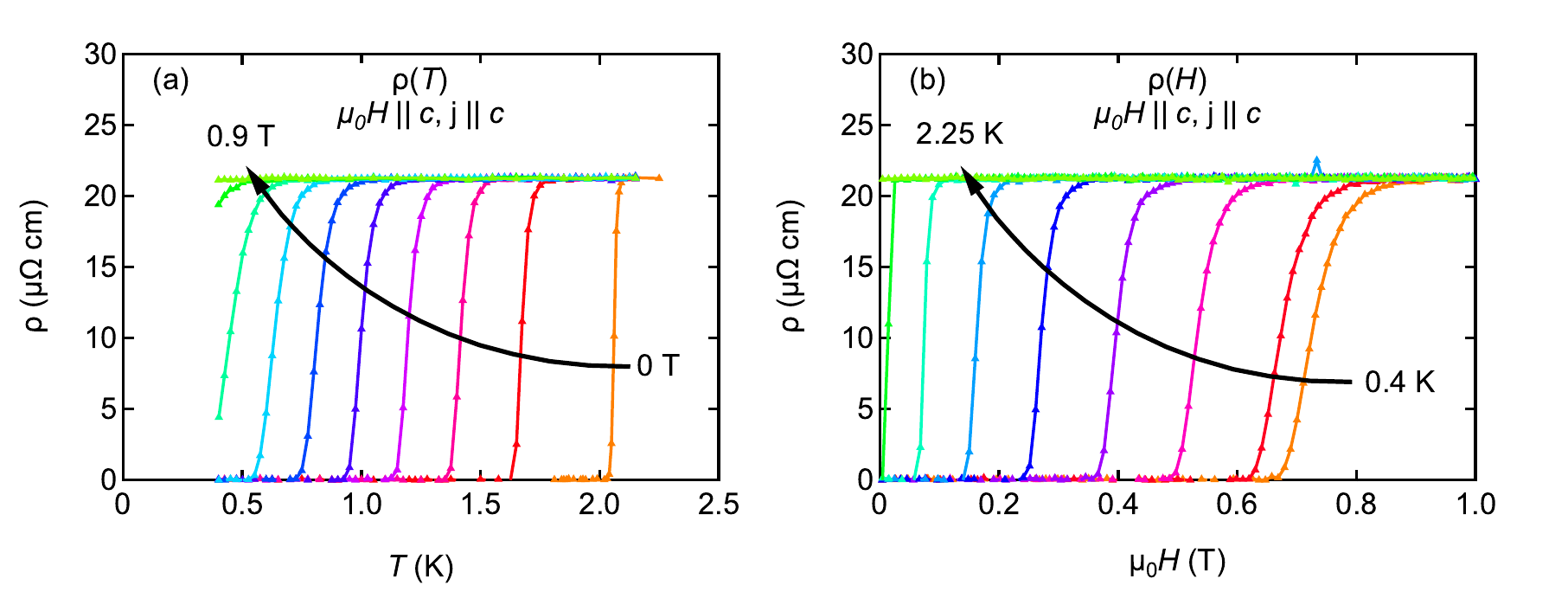}
		\caption{Low temperature resistivity data showing the superconducting transition with $\rho(T)$ (a) and $\rho(H)$ (b) sweeps. The $\rho(T)$ sweeps have a constant magnetic field $\mu_0H\,=0-0.9$\,T in increments of $0.1$\,T. The $\rho(H)$ sweeps have a constant temperature $T\,=0.5-2.25$\,K and $0.4$\,K. These representative measurements were complected with $H\parallel\,c$ and j\,$\parallel\,c$.}
		\label{fig:LowTempRes}
\end{figure}

Resistivity measurements under field were conducted to construct the anisotropic upper-critical-field phase diagram. Measurements were completed by performing three sets of temperature and field sweeps. Each set had the external magnetic field aligned along a different crystallographic axis. Fig.~\ref{fig:LowTempRes} shows the resistivity data when the magnetic field was aligned parallel to the $c$-axis.

\section{Heat Capacity}

\begin{figure}[t]
	\centering
	\includegraphics[width=\textwidth]{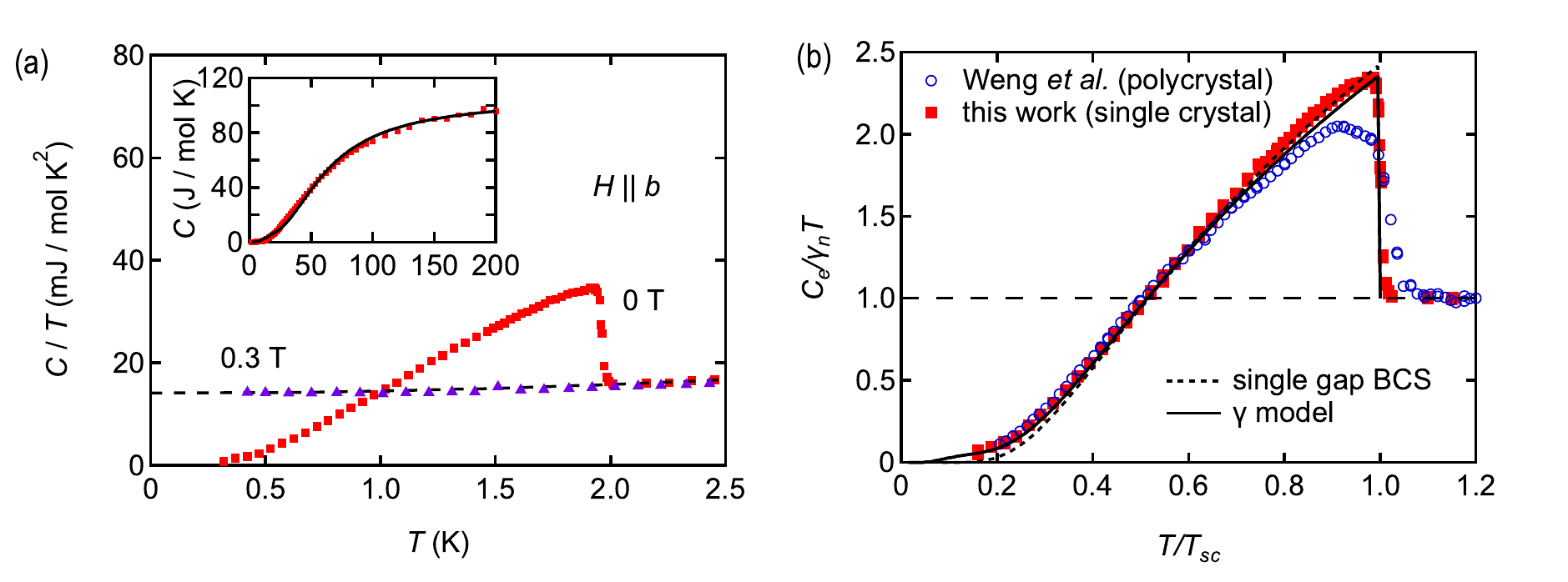}
	\caption{(a) Zero-field specific heat ($C/T$) against $T$ shows the complete superconducting transition. The purple curve shows that a $0.3$\,T external field is sufficient to suppress superconductivity below $0.4$\,K in heat capacity measurements, a lower value than in resistivity measurements. The dashed black line is the low-temperature $T^{2}$ phonon contribution. The inset shows the high-temperature heat capacity, which is fitted to the weighted high-temperature Einstein-Debye model, Eq.\ref{eq:HeatCapacity}. (b) Comparison of the electronic heat capacity measured on polycrystal~\cite{Weng2016} with our measurement on single crystal. The single gap BCS and a two-gap model based on Ref.~\cite{Kogan2009PRB} ($\gamma$-model) are shown. The parameters used for the $\gamma$-model are $n_1=0.95$, $\lambda_{12}=0.1$, $\lambda_{11}=\lambda_{22}=0.45$.}
	\label{fig:Thermoab}
\end{figure}

Fig.~\ref{fig:Thermoab}(a) shows the zero-field heat capacity across a range of $0.4$-$200$\,K. The low-temperature specific heat ($C/T$) shows a linear relationship versus $T^{2}$ and is fit according to the formula $C/T=\gamma_{n}+\beta\,T^{2}$. From the fit, a Debye temperature $\Theta_D^L=166$\,K and a Sommerfeld coefficient $\gamma_{n}=14.1$\,mJ/mol\,K$^2$ are obtained. These values indicate that LaNiGa$_2$ does not exhibit strong electronic correlations, as expected for a La-based material. Additionally, the high-temperature data is well fit to a weighted high-temperature Einstein-Debye model \cite{Gopal1966}:

\begin{subequations}
	\begin{align}
		\label{eq:HeatCapacity}
		C&(T) = \gamma_{n}T + n x C_\textrm{Debye}\left(\frac{T}{\Theta_\textrm{D}}\right) + n(1-x)C_\textrm{Einstein}\left(\frac{T}{\Theta_{E}}\right) \\
		C&_\textrm{Debye}\left(\frac{T}{\Theta_{D}}\right) = 9R\left(\frac{T}{\Theta_{D}}\right)^{3} \int_{0}^{\frac{\Theta_{D}}{T}} \frac{x^{4}e^{x}}{(e^{x} - 1)^{2}} dx \\
		C&_\textrm{Einstein}\left(\frac{T}{\Theta_{E}}\right) = 3R\frac{z^{2}e^{z}}{(e^{z} - 1)^{2}}, z = \frac{\Theta_{E}}{T}
	\end{align}
\end{subequations}

where $n$ is the number of atoms in a formula unit and $x$ is the fractional contribution of Debye model. In both models, $C_\textrm{Debye}$ and $C_\textrm{Einstein}$, there is a single refineable parameter of $\Theta_D$ and $\Theta_{E}$, respectively. Since the Einstein model is used to approximate the optical phonon contributions \cite{Kittel2005}, it is best to calculate the total number of phonons branches to better estimate the weighted contribution of each heat capacity model. With the new structure, the primitive cell volume of LaNiGa$_2$ is half that of the unit cell, thus $8$ atoms in the primitive cell. It follows that there are $24$ phonon branches, three of which are acoustic and will have strong contributions from the heavy La atom. The effective Debye model contribution should be $x=12.5\%$. The inset of Fig.~\ref{fig:Thermoab}(a) shows the fitting of the function which gives a Debye temperature, $\Theta_D=83$\,K, and an Einstein temperature $\Theta_{E}=200$\,K.

A complete bulk superconducting transition is observed. The midpoint of the transition is $T_{sc}^C=1.96$\,K. When normalized with the $\gamma_{n}$ value from the low-temperature fit, this specific heat jump equates to $\Delta\,C/\gamma_{n}T_{sc}^C=1.33$, slightly higher than previously reported value from \cite{Weng2016PRL} on polycrystals (see Fig.~\ref{fig:Thermoab}(b) for comparison). Though the specific heat near $T_{sc}$ seems to be well described by the single-gap BCS theory, the low temperature data can be better described by a two-gap model~\cite{Kogan2009PRB} (see Fig.~\ref{fig:Thermoab}(b)) as already reported for polycrystals~\cite{Weng2016PRL}. We note that the heat capacity is reported down to $0.4$\,K, which is significantly higher than the reported penetration depth measurements down to $0.05$\,K upon which the nodeless multigap behavior was inferred~\cite{Weng2016PRL}. Heat capacity measurements at lower temperatures are necessary to better assess the superconducting gap structure.

The Kadowaki-Woods ratio (KWR)\cite{Kadowaki1986,Jacko2009} calculated as $A/\gamma^{2}_{n}$ is equal to $1.28$\,$\mu\Omega$\,cm\,mol$^{2}$\,K$^{2}$\,J$^{-2}$ confirming that LaNiGa$_2$ is not a strongly correlated material.

Using both the normal and superconducting-state heat capacity data, the isotropic thermodynamic critical field, $H_{c}(T)$, can be calculated using:

\begin{equation}
	\frac{\mu_{0} V_{m} H_{c} (T)^{2}}{2} = \int_{T}^{T_{sc}} \Delta S(x) dx
\end{equation}

where $\Delta S(T)$ is the entropy difference between the normal and superconducting states. $V_{m}$ is the molar volume from the new crystal structure~\cite{Annett2004}. By fitting the $H_{c}(T)$ curve with a Taylor expansion fit, a value of $23$\,mT is obtained for $\mu_{0}H_{c}(0)$~\cite{Decker1958}.

\begin{figure}
	\centering
	\includegraphics{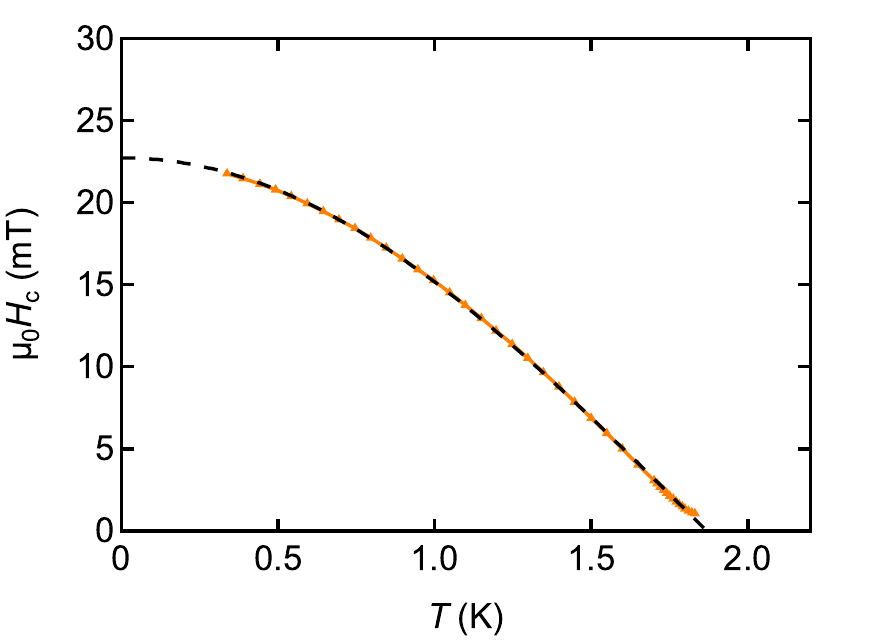}
	\caption{Temperature dependent thermodynamic critical field, $\mu_{0}H_{c}(T)$, data that was calculated from the entropy difference, $\Delta S(T)$, of the normal and superconducting-state heat capacity data.}
	\label{fig:CriticalField}
\end{figure}

\section{Superconducting Phase Diagram}

\begin{figure}[!htb]
    \centering
	\includegraphics{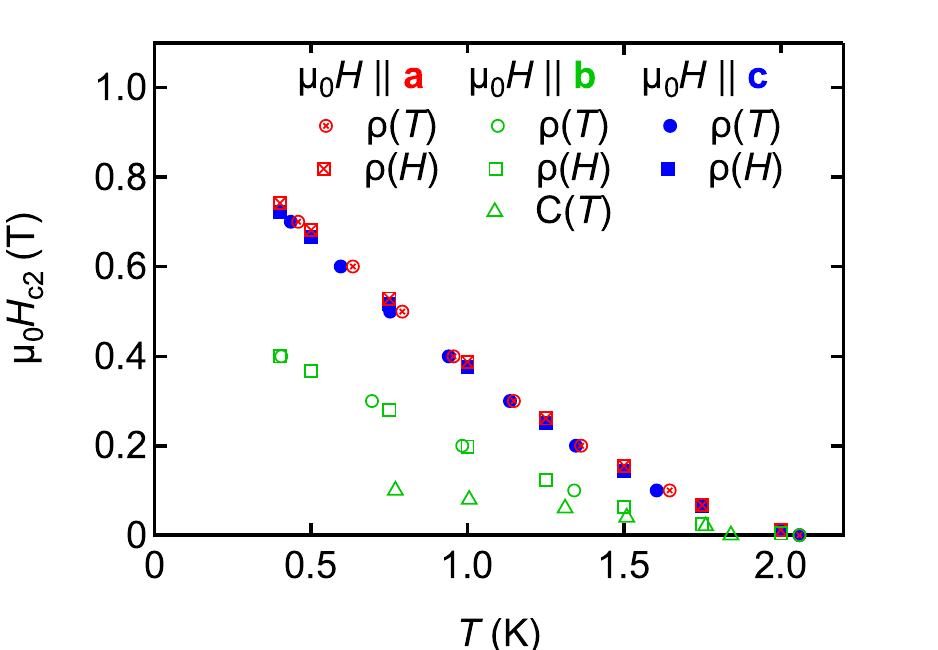}
	\caption{Anisotropic field-temperature $\mu_{0}H_{c2}$ phase diagrams of LaNiGa$_2$ when $H$ is applied along each of the three crystallographic axes.}
	\label{fig:PhaseDiagram}
\end{figure}

The anisotropic $H_{c2}$ phase diagram is constructed by tracking the superconducting transition across resistivity and heat capacity temperature- and field-sweeps for an aligned LaNiGa$_2$ crystal. These measurements were collected with an external magnetic field carefully orientated along particular crystallographic axes (Fig.~\ref{fig:PhaseDiagram}). Regardless of the orientation, there is a clear upward curvature of the $H_{c2}$, which is a common feature for multiband superconductivity and was previously noted on polycrystalline samples of LaNiGa$_2$\cite{Weng2016}. 

Additionally, from the Helfand-Werthamer model the critical field at $0$\,K can be approximated for a single band system: 
\begin{equation}
    \mu_{0}H_{c2}(0) = -AT_{sc}\frac{d\mu_{0}H_{c2}(T_{sc})}{dT},
\end{equation}

where $A=0.73$ and $0.69$ for the clean and dirty limits, respectively \cite{Helfand1966}. Estimated values of $H_{c2}(0)$ from the slope near $T_{sc}$ are lower than $0.275$\,T for all three field directions, and lower than the experimental values near $0.4$\,K. Thus indicating that a single band model, either in the clean or dirty limit, cannot accurately model this system. This is further evidence for the suggestion that multi-band effects are important. The experimental and calculated superconducting properties are summarized in Table~\ref{tab:Properties}. Additionally, from the critical temperature, $T_{sc}^C$, the Pauli paramagnetic limit is calculated by $\mu_{0}^{p}H_{c2}(0)=1.84T_{sc}^C=3.66$\,T \cite{Clogston1962}. Since all three axes have critical fields below this limit, an orbital pair-breaking mechanism may be operating. As discussed in the main text, LaNiGa$_2$ is not a heavy fermion material ($\gamma_{n}=14.1$\,mJ mol$^{-1}$ K$^{-2}$) and interband pairing is suppressed by the application of a strong magnetic field. Thus the upper-critical field in LaNiGa$_2$ remains orbital limited, even though time-reversal symmetry breaking superconductivity was observed at zero field in $\mu$SR experiments.

\begin{table}[!htb]
	\centering
		\begin{tabular}{ l c c }
		\hline
		\hline
			Property (Unit) & Value  & Previously Reported\\
		\hline
			$T_{sc}$ (K) & $1.96$ & $1.8$~\cite{Weng2016}\\
			$\gamma_{n}$ (mJ\,mol$^{-1}$\,K$^{-2}$) & $14.1$ & $10.54$~\cite{Weng2016}\\
			$\Theta_D$ (low temp.) (K) & $166$ & $294$~\cite{Weng2016}\\
			$\Theta_D$ (K) & $82.7$ & -\\
			$\Theta_{E}$\ (K) & $200$ & -\\
			$\Delta\,C/\gamma_{n}T_{sc}$ & $1.33$ & $1.28$~\cite{Weng2016}\\
			$\rho_{0}$ ($\mu\Omega$\,cm) & $5.20$ & $1.6$~\cite{Weng2016}\\
			A ($\mu\Omega$\,cm\,T$^{-2}$) & $2.54\times10^{-4}$ & - \\
			$\mu_{0}H^{HW}_{c2}$($0$) (Clean limit) (T) & $0.275$, $0.094$, and $0.253$ & $\approx\,0.06^{\dagger}$~\cite{Weng2016}\\
			$\mu_{0}H^{HW}_{c2}$($0$) (Dirty limit) (T) & $0.260$, $0.089$, and $0.239$ & -\\
			$\mu_{0}H_{c2}^{p}$($0$) (T) & $3.66$ & -\\
			$\mu_{0}H_{c}$($0$) (mT) & $23$ & -\\
			$\xi_{GL}$ (nm) & $51.5$, $17.6$, and $47.3$ & $28$*~\cite{Hillier2012}\\
			$\lambda_{GL}$ (nm) & $174$, $509$, and $189$ & $350$*~\cite{Hillier2012}\\
			$\kappa$ & $3.38$, $28.9$, $4.00$ & $12.5$*~\cite{Hillier2012}\\
			KWR = A/$\gamma_{n}^{2}$ ($\mu\Omega$\,cm\,mol$^{2}$\,K$^{2}$\,J$^{-2}$) & $1.28$ & -\\
		\hline
		\hline			
		\end{tabular}
	\caption{Measured and calculated relevant normal and superconducting state properties for LaNiGa$_2$. All anisotropic parameters have entries in the following order: $a$-, $b$-, and $c$-axis. The $T_{sc}$ was selected as the midpoint from the heat capacity transition. $^{\dagger}$Not specified whether clean or dirty limity. *Calculated from field dependence of the muon depolarization rate on a polycrystalline sample.}
	\label{tab:Properties}
\end{table} 

\section{Superconducting and Physical Properties}

Within the superconducting state, we can calculate the anisotropic Ginzburg-Landau (GL) coherence lengths, $\xi_{GL}$, by using the relation: 

\begin{equation}
	\frac{d(\mu_{0}H^{a}_{c2}(T_{sc}))}{dT} = \frac{-\Phi_{0}}{2\pi\xi^{b}_{GL}\xi^{c}_{GL}T_{sc}}
\end{equation}

 where $\Phi_{0}$ is the quantum flux, $\mu_{0}H^{a}_{c2}$ is the of the $\mu_{0}H_{c2}$ curve when field is parallel to the crystallographic $a$-axis, and $\xi^{b}_{GL}$ and$\xi^{c}_{GL}$ are the coherence lengths along the $b$- and $c$-axes \cite{TinkhamIntroSupra,Annett2004}. Given the orthorhombic nature, by measuring the slope of $H_{c2}$ along each axis near $T_{sc}$ we can find the corresponding coherence lengths. From this linear system of equations, $\xi^{a}_{GL}$, $\xi^{b}_{GL}$, $\xi^{c}_{GL}$ are calculated out to $51.5$, $17.6$, and $47.3$\,nm, respectively.

With the $H_{c}$($0$) and the anisotropic $\xi_{GL}$ values, the anisotropic penetration depths can be calculated with \cite{Annett2004,TinkhamIntroSupra}:

\[
\mu_{0}H_{c}(0) = \frac{\Phi_{0}}{2\sqrt{2}\pi\xi^{a}_{GL}\lambda^{a}_{GL}} 
\]

From this $\lambda_{GL}$ for each crystallographic axis is calculated to $174$, $509$, and $189$\,nm for the $a$-, $b$-, and $c$-axes, respectively. When averaged across the three penetration depths $\lambda_{GL}^{avg}=291$\,nm, which is in great agreement with the previoulsy reported penetration depth $\lambda_{0}=350$\,nm~\cite{Weng2016}. Lastly, with $\lambda_{GL}$ and $\xi_{GL}$, the $\kappa$ ratio can be determined along each axis: $\kappa^{a}$ = $3.37$, $\kappa^{b}$ = $28.9$, and $\kappa^{c}$ = $4.00$. 

\section{Nonsymmorphic Symmetry Analysis}

LaNiGa$_2$ crystallizes in the orthorhombic structure (space group nb.~63 \textit{Cmcm}).
This space group can be generated by the symmetry operations (ignoring spin):

\begin{tabular}{lll}
identity&$\mathcal{I}$&$(x,y,z)\rightarrow(x,y,z)$\\
inversion&$\mathcal{P}$&$(x,y,z)\rightarrow(-x,-y,-z)$\\
2-fold screw axis&$\mathcal{S}_{2}^z$&$(x,y,z)\rightarrow(-x,-y,z+\frac{1}{2})$\\
2-fold rotation&$\mathcal{C}_{2}^y$&$(x,y,z)\rightarrow(-x,y,-z+\frac{1}{2})$
\end{tabular}

from which we can also obtain the additional symmetries~\cite{IntTablesCrystallo}:

\begin{tabular}{lll}
2-fold rotation&$\mathcal{C}_{2}^x$&$(x,y,z)\rightarrow(x,-y,-z)$\\
reflection&$\mathcal{M}_z$&$(x,y,z)\rightarrow(x,y,-z+\frac{1}{2})$\\
glide&$\mathcal{G}_y$&$(x,y,z)\rightarrow(x,-y,z+\frac{1}{2})$\\
reflection&$\mathcal{M}_x$&$(x,y,z)\rightarrow(-x,y,z)$
\end{tabular}

We can combine these symmetry operations to obtain in particular:
\begin{eqnarray*}
\mathcal{P}\mathcal{S}_{2}^z: (x,y,z)&\rightarrow&(x,y,-z-\frac{1}{2})\\
(\mathcal{P}\mathcal{S}_{2}^z)^2: (x,y,z)&\rightarrow&(x,y,z)
\end{eqnarray*}
In momentum space, $\mathcal{P}$ reverses $k$, and $\mathcal{S}_{2}^z$ reverses $k_x$ and $k_y$ but preserves $k_z$~\cite{Wu2018PRB}. So the combined operation $\mathcal{P}\mathcal{S}_{2}^z$ only reverses the $k_z$ component. This implies that the planes $k_z=0$ and $k_z=\dfrac{\pi}{c}$ are in the invariant subspace of $\mathcal{P}\mathcal{S}_{2}^z$. Since $(\mathcal{P}\mathcal{S}_{2}^z)^2=1$, the eigenvalues of $\mathcal{P}\mathcal{S}_{2}^z$ on the $k_z=0$ and $k_z=\dfrac{\pi}{c}$ planes are $\pm1$. So we can write: $\mathcal{P}\mathcal{S}_2^z|\psi_\pm\rangle=\pm |\psi_\pm\rangle$.

In addition, we have $\mathcal{P}\mathcal{T}$ symmetry everywhere in the Brillouin zone, and the commutation relation between $\mathcal{P}\mathcal{T}$ and $\mathcal{P}\mathcal{S}_{2}^z$ is:
\begin{eqnarray*}
\mathcal{P}\mathcal{T}\mathcal{P}\mathcal{S}_{2}^z=T_{001}\mathcal{P}\mathcal{S}_{2}^z\mathcal{P}\mathcal{T}=e^{-ik_zc}\mathcal{P}\mathcal{S}_{2}^z\mathcal{P}\mathcal{T}
\end{eqnarray*}
Therefore, on the $k_z=\dfrac{\pi}{c}$ plane, we have the anticommutation $\left\{\mathcal{P}\mathcal{T},\mathcal{P}\mathcal{S}_{2}^z\right\}=0$ which gives:
\begin{eqnarray*}
\mathcal{P}\mathcal{S}_2^z\mathcal{P}\mathcal{T}|\psi_\pm\rangle=\mp\mathcal{P}\mathcal{T}|\psi_\pm\rangle
\end{eqnarray*}
So $|\psi_\pm\rangle$ and $\mathcal{P}\mathcal{T}|\psi_\pm\rangle$ have opposite eigenvalues, imposing that each band is 2-fold degenerate~\cite{Young2015PRL}. Adding the spin degrees of freedom, we have 4-fold degeneracy at the $k_z=\dfrac{\pi}{c}$ plane. Interestingly, if time reversal symmetry is broken in the spin degrees of freedom and without SOC, one still has a 2-fold degeneracy for each spin species at the $k_z=\dfrac{\pi}{c}$ plane~\cite{Wu2018PRB}.

In the presence of SOC, we still have 2-fold degeneracy because $(\mathcal{P}\mathcal{T})^2=-1$, implying that $|\psi_\pm\rangle$ and $\mathcal{P}\mathcal{T}|\psi_\pm\rangle$ have the same eigenvalues. We now have to take into account the effect of the symmetry operations on the spin space. The square of $\mathcal{S}_2^z$ will rotate the spin by $2\pi$, so that:
\begin{eqnarray*}
(\mathcal{P}\mathcal{S}_{2}^z)^2: (x,y,z)&\rightarrow&(x,y,z)\times \overline{E}
\end{eqnarray*}
where $\overline{E}$ is a $2\pi$ rotation of spins. Since $(\mathcal{P}\mathcal{S}_{2}^z)^2$=$-1$, the eigenvalues of $\mathcal{P}\mathcal{S}_{2}^z$ on the $k_z=0$ and $k_z=\dfrac{\pi}{c}$ planes are $\pm i$. So we can write: $\mathcal{P}\mathcal{S}_2^z|\psi_\pm\rangle=\pm i |\psi_\pm\rangle$. On the $k_z=\dfrac{\pi}{c}$ plane, we still have the anticommutation $\left\{\mathcal{P}\mathcal{T},\mathcal{P}\mathcal{S}_{2}^z\right\}=0$ which gives:
\begin{eqnarray*}
\mathcal{P}\mathcal{S}_2^z\mathcal{P}\mathcal{T}|\psi_\pm\rangle=\pm i\mathcal{P}\mathcal{T}|\psi_\pm\rangle
\end{eqnarray*}
So $|\psi_\pm\rangle$ and $\mathcal{P}\mathcal{T}|\psi_\pm\rangle$ have the same eigenvalues, and we obtain 2 Kramers pairs~\cite{Liang2016PRB} ($|\psi_+\rangle$,$\mathcal{P}\mathcal{T}|\psi_+\rangle$ and $|\psi_-\rangle$,$\mathcal{P}\mathcal{T}|\psi_-\rangle$).

However, the 2 eigenstates of $\mathcal{P}\mathcal{S}_{2}^z$ can be related by additional symmetries. For example, the $Z$-$T$ line, corresponding to (0,$k_y,\dfrac{\pi}{c})$, is invariant under $\mathcal{M}_x$. The anticommutation relation $\left\{\mathcal{P}\mathcal{S}_{2}^z,\mathcal{M}_x\right\}=0$ gives:
\begin{eqnarray*}
\mathcal{P}\mathcal{S}_{2}^z\mathcal{M}_x|\psi_\pm\rangle=\mp i\mathcal{M}_x|\psi_\pm\rangle
\end{eqnarray*}
So $|\psi_\pm\rangle$ and $\mathcal{M}_x|\psi_\pm\rangle$ have opposite eigenvalues which guarantees a 4-fold degeneracy: $|\psi_\pm\rangle$, $\mathcal{M}_x|\psi_\pm\rangle$, $\mathcal{P}\mathcal{T}|\psi_\pm\rangle$, and $\mathcal{P}\mathcal{T}\mathcal{M}_x|\psi_\pm\rangle$~\cite{Liang2016PRB}. The degeneracy leads to the true-Dirac points represented in Fig.~2(c) (main text).

We can check that this degeneracy is not guaranteed on the $Z$-$A$ line, corresponding to $(k_x,0,\dfrac{\pi}{c})$, invariant under $\mathcal{G}_y$. The commutation relation is:
\begin{eqnarray*}
\mathcal{P}\mathcal{S}_{2}^z\mathcal{G}_y=-T_{00-1}\mathcal{G}_y\mathcal{P}\mathcal{S}_{2}^z
\end{eqnarray*}
which gives, on the $k_z=\dfrac{\pi}{c}$ plane, the commutation $\left[\mathcal{P}\mathcal{S}_{2}^z,\mathcal{G}_y\right]=0$. Then:
\begin{eqnarray*}
\mathcal{P}\mathcal{S}_{2}^z\mathcal{G}_y|\psi_\pm\rangle=\pm i\mathcal{G}_y|\psi_\pm\rangle
\end{eqnarray*}
We see that $|\psi_\pm\rangle$ and $\mathcal{G}_y|\psi_\pm\rangle$ have the same eigenvalues, so $\mathcal{G}$ does not relate the two eigenstates. On $Z$-$A$, we remain with two Kramers pairs.

\section{Gap Structure and Symmetry}

We now discuss the pairing gap structures allowed by symmetry in the space group $Cmcm$ (\#63) with point group $D_{2h}$. The basis functions and nodes inside the Brillouin zone (BZ) are taken from Annett's compilation.~\cite{Annett1990AP} With the use of modern group theoretical and topological classification theory of superconducting nodes~\cite{Yarzhemsky1992JPCM,Yarzhemsky1998PSSB,Yarzhemsky2000IJQC,Yarzhemsky2003AIPCP,Yarzhemsky2008JOE,Micklitz2009PRB,Micklitz2017PRB,Micklitz2017PRL,Nomoto2017JPSJ,Sumita2018PRB,Kobayashi2018PRB,Sumita2019PRB,Sumita2021PhD,Ono2021}, \\
$\bullet$ the nodes at a general position on the $k_z=\pi/c$ plane, \\
$\bullet$ on the ($0$,$k_y$,$\pi/c$) line ($Z$-$T$ line for $a<b$, $Z$-$B$ line for $a>b$), and \\
$\bullet$ on the ($k_x$,$0$,$\pi/c$) line ($Z$-$A$ line for $a<b$, $Z$-$T$ line for $a>b$) \\
are also indicated below. In the case with weak SOC, interband pairing is possible on the $k_z$=$\pi/c$ plane. The product of the pure orbital function with the sublattice function can be symmetric (s.) or antisymmetric (a.s.) under fermion exchange. The symmetric case corresponds to spin-singlet, and the antisymmetric case corresponds to spin-triplet. Green or blue colors indicate the additional possibilities due to interband pairing. Blue color indicates that time-reversal symmetry can be broken in the spin-channel. The nodes on the $k_z=\pi/c$ plane for strong SOC are taken from~\cite{Sumita2021PhD,Sumita2018PRB}. Green or blue colors indicate a difference with a symmorphic space group such as $Cmmm$ (\#65). Blank cells indicate the structure has not yet been determined.\\

 In the case with weak spin-orbit coupling:\\
\begin{small}
\begin{tabular}{|l|l|l|l|l|l|l|l|l|l|l|} \hline
\multirow{3}{1mm}{state} & \multirow{3}{15mm}{$d/\Delta_0$ inside BZ} & \multirow{3}{15mm}{$\Delta/\Delta_0$ inside BZ} & \multirow{3}{14mm}{unitary? inside BZ} & \multicolumn{7}{c|}{nodes}\\ \cline{5-11}
 & & & & \multirow{2}{15mm}{inside BZ} & \multicolumn{2}{c|}{($k_x$,$k_y$,$\pi/c$)} & \multicolumn{2}{c|}{($0$,$k_y$,$\pi/c$)} & \multicolumn{2}{c|}{($k_x$,$0$,$\pi/c$)}\\ \cline{6-11}
 & & & & & s. & a.s. & s. & a.s. & s. & a.s.\\ \hline \hline
$^1A_{1g}$ & singlet & $1\left(|\uparrow\downarrow\rangle-|\downarrow\uparrow\rangle\right)$ & singlet & fully gapped & gap &\cellcolor{cyan}gap & gap &\cellcolor{cyan}gap & gap &node\\ \hline
$^1B_{1g}$ & singlet & $|XY|\left(|\uparrow\downarrow\rangle-|\downarrow\uparrow\rangle\right)$ & singlet & lines at $k_x, k_y=0$ & gap &\cellcolor{cyan}gap & node & node & node &\cellcolor{cyan}gap\\ \hline
$^1B_{2g}$ & singlet & $|XZ|\left(|\uparrow\downarrow\rangle-|\downarrow\uparrow\rangle\right)$ & singlet & lines at $k_x, k_z=0$ &\cellcolor{green}gap&node & node &node & \cellcolor{green}gap &node\\ \hline
$^1B_{3g}$ & singlet & $|YZ|\left(|\uparrow\downarrow\rangle-|\downarrow\uparrow\rangle\right)$ & singlet & lines at $k_y, k_z=0$ &\cellcolor{green}gap&node & \cellcolor{green}gap &node & \cellcolor{green}gap &node\\ \hline
$^3A_{1ua}$ & $(0,0,1)XYZ$ & $1|XYZ| \left(|\uparrow\downarrow\rangle+|\downarrow\uparrow\rangle\right)$ & unitary & lines at $k_x, k_y, k_z=0$ &node&\cellcolor{green}gap &node & node &node &\cellcolor{green}gap\\ \hline
$^3A_{1ub}$ & $(1,i,0)XYZ$ & \begin{tabular}{l} $2|XYZ||\uparrow\uparrow\rangle$\\ $0|\downarrow\downarrow\rangle$\end{tabular} & non-unitary & \begin{tabular}{l} lines at $k_x, k_y, k_z=0$\\ surface \end{tabular} & N/A & \begin{tabular}{l} ~ \\ ~ \end{tabular} & N/A & & N/A & \\ \hline
$^3B_{1ua}$ & $(0,0,1)Z$ & $1|Z| \left(|\uparrow\downarrow\rangle+|\downarrow\uparrow\rangle\right)$ & unitary & line at $k_z=0$ &node&\cellcolor{green}gap &node &\cellcolor{green}gap &node &\cellcolor{green}gap\\ \hline
$^3B_{1ub}$ & $(1,i,0)Z$ & \begin{tabular}{l} $2|Z| |\uparrow\uparrow\rangle$\\ $0|\downarrow\downarrow\rangle$\end{tabular} & non-unitary & \begin{tabular}{l} line at $k_z=0$\\ surface \end{tabular} & N/A & \begin{tabular}{l} ~ \\ ~ \end{tabular} & N/A & & N/A & \\ \hline
$^3B_{2ua}$ & $(0,0,1)Y$ & $1|Y| \left(|\uparrow\downarrow\rangle+|\downarrow\uparrow\rangle\right)$ & unitary & line at $k_y=0$ &\cellcolor{green}gap & gap &\cellcolor{green}gap & gap &\cellcolor{green}gap & node \\ \hline
$^3B_{2ub}$ & $(1,i,0)Y$ & \begin{tabular}{l} $2|Y| |\uparrow\uparrow\rangle$\\ $0|\downarrow\downarrow\rangle$\end{tabular} & non-unitary & \begin{tabular}{l} line at $k_y=0$\\ surface \end{tabular} & N/A & \begin{tabular}{l} ~ \\ ~ \end{tabular} & N/A & & N/A & \\ \hline
$^3B_{3ua}$ & $(0,0,1)X$ & $1|X| \left(|\uparrow\downarrow\rangle+|\downarrow\uparrow\rangle\right)$ & unitary & line at $k_x=0$ &\cellcolor{green}gap & gap & node & node & node & gap\\ \hline
$^3B_{3ub}$ & $(1,i,0)X$ & \begin{tabular}{l} $2|X| |\uparrow\uparrow\rangle$\\ $0|\downarrow\downarrow\rangle$\end{tabular} & non-unitary & \begin{tabular}{l} line at $k_x=0$\\ surface \end{tabular} & N/A & \begin{tabular}{l} ~ \\ ~ \end{tabular} & N/A & & N/A &\\ \hline
\end{tabular}
\end{small}
~\\
~\\

In the case with strong spin-orbit coupling:\\
\begin{small}
\begin{tabular}{|l|l|l|l|l|l|l|l|} \hline
\multirow{2}{7mm}{state} & \multirow{2}{15mm}{$d/\Delta_0$ inside BZ} & \multirow{2}{15mm}{$\Delta/\Delta_0$ inside BZ} & \multirow{2}{14mm}{unitary? inside BZ} & \multicolumn{4}{c|}{nodes}\\ \cline{5-8}
 & & & & inside BZ & ($k_x$,$k_y$,$\pi/c$) & ($0$,$k_y$,$\pi/c$) & ($k_x$,$0$,$\pi/c$)\\ \hline \hline
$A_{1g}$ & singlet & $1\left(|\uparrow\downarrow\rangle-|\downarrow\uparrow\rangle\right)$ & singlet & fully gapped & gap & gap & gap\\ \hline
$B_{1g}$ & singlet & $|XY|\left(|\uparrow\downarrow\rangle-|\downarrow\uparrow\rangle\right)$ & singlet & lines at $k_x, k_y=0$ & gap &\cellcolor{green}gap & node\\ \hline
$B_{2g}$ & singlet & $|XZ|\left(|\uparrow\downarrow\rangle-|\downarrow\uparrow\rangle\right)$ & singlet & lines at $k_x, k_z=0$ & node &\cellcolor{green}gap & node\\ \hline
$B_{3g}$ & singlet & $|YZ|\left(|\uparrow\downarrow\rangle-|\downarrow\uparrow\rangle\right)$ & singlet & lines at $k_y, k_z=0$ & node &\cellcolor{green}gap & node\\ \hline
$A_{1u}$ & $(AX,BY,CZ)$ & \begin{tabular}{l}$\sqrt{A^2X^2-B^2Y^2} |\uparrow\uparrow\rangle$\\$+|CZ|\left(|\uparrow\downarrow\rangle+|\downarrow\uparrow\rangle\right)$\\$+\sqrt{A^2X^2-B^2Y^2} |\downarrow\downarrow\rangle$ \end{tabular} & unitary & none &\cellcolor{green}node & gap &\cellcolor{green}node\\ \hline
$B_{1u}$ & $(AY,BX,CXYZ)$ & \begin{tabular}{l}$\sqrt{A^2Y^2-B^2X^2} |\uparrow\uparrow\rangle$\\$+|CXYZ|\left(|\uparrow\downarrow\rangle+|\downarrow\uparrow\rangle\right)$\\$+\sqrt{A^2Y^2-B^2X^2} |\downarrow\downarrow\rangle$ \end{tabular} & unitary & point on $k_z$ axis &\cellcolor{green}node & gap &\cellcolor{green}node\\ \hline
$B_{2u}$ & $(AZ,BXYZ,CX)$ & \begin{tabular}{l}$\sqrt{A^2Z^2-B^2X^2Y^2Z^2} |\uparrow\uparrow\rangle$\\$+|CX|\left(|\uparrow\downarrow\rangle+|\downarrow\uparrow\rangle\right)$\\$+\sqrt{A^2Z^2-B^2X^2Y^2Z^2} |\downarrow\downarrow\rangle$ \end{tabular} & unitary & point on $k_y$ axis & gap & gap & gap\\ \hline
$B_{3u}$ & $(AXYZ,BZ,CY)$ & \begin{tabular}{l}$\sqrt{A^2X^2Y^2Z^2-B^2Z^2} |\uparrow\uparrow\rangle$\\$+|CY|\left(|\uparrow\downarrow\rangle+|\downarrow\uparrow\rangle\right)$\\$+\sqrt{A^2X^2Y^2Z^2-B^2Z^2} |\downarrow\downarrow\rangle$ \end{tabular} & unitary & point on $k_x$ axis & gap & gap & gap\\ \hline
\end{tabular}
\end{small}
\vskip 3mm

Strong SOC means splitting of bands on the Fermi surface larger than the superconducting gap $2\Delta_0$, weak SOC means the alternative regime. LaNiGa$_2$ presents an interesting complication. The superconducting gap is roughly $3k_BT_c\approx\,0.5$\,meV. This is well below SOC splitting of crossings (or movements of bands) by a factor of 10-50 over most of the Fermi surface. However, as mentioned, the failure of SOC to split bands along the $Z$-$T$ line on the Brillouin zone face leads to two Dirac points outlasting SOC in the normal state, fixed at the Fermi surface independent of chemical potential variations, giving rise to an unanticipated type of topological superconductivity with remaining BdG degeneracies in the superconducting state.

The point here is that, at these diabolical points, the SOC splitting vanishes. Thus in a region around these points, LaNiGa$_2$ is in the weak SOC regime, and this is where, or very nearly where, the minimum superconducting gap occurs. Some of the consequences for the BdG quasiparticle dispersion have been presented in the main text.

\section{ARPES}

\begin{figure}[!ht]
    \centering
	\includegraphics[width=\columnwidth]{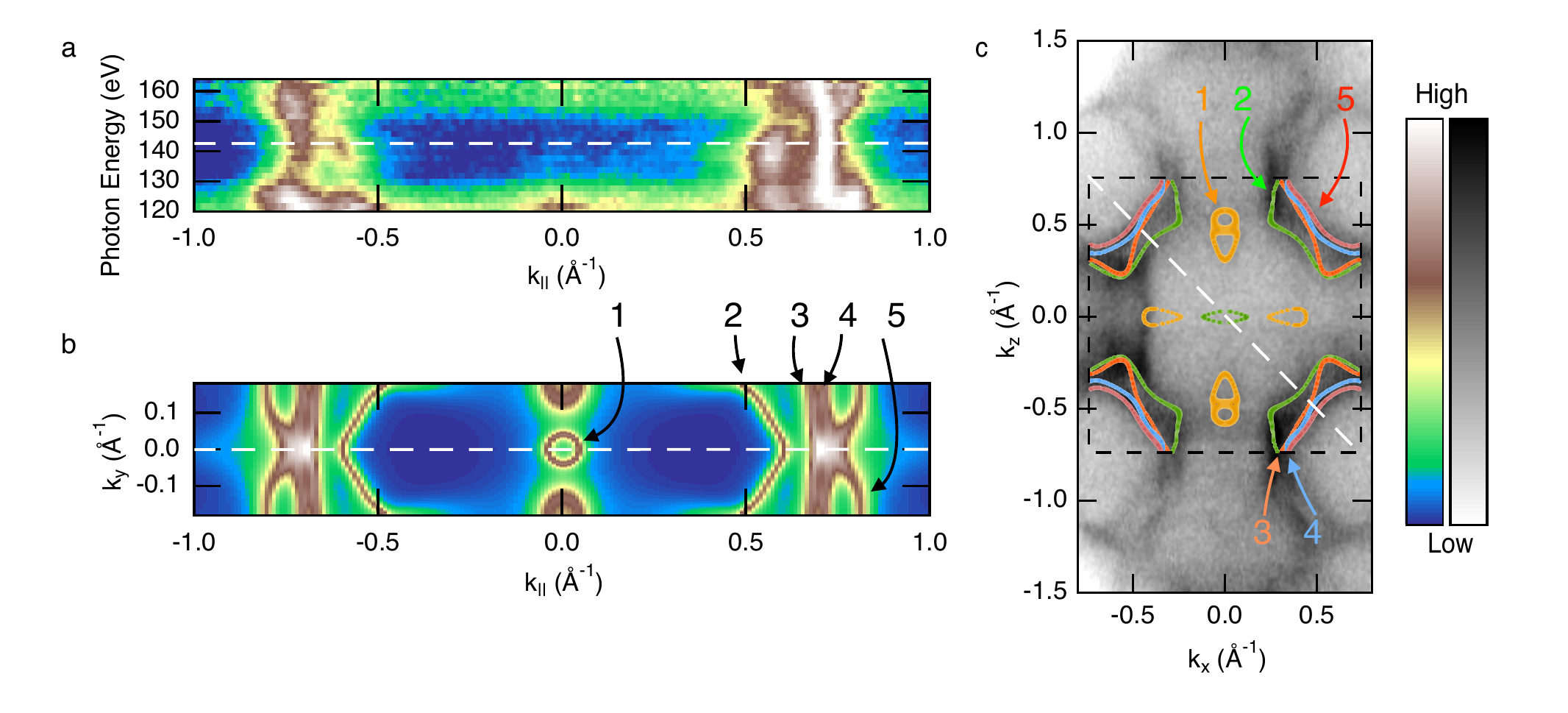}
	\caption{Comparison of ARPES to DFT. (a) Photon energy dependence of ARPES intensity along the $\Gamma$-$A$ diagonal, $k_{||}$, through the BZ. (b) DFT calculation of spectral weight $A(\vec{k},\omega)$ along the same plane as (a). Comparing the structure of (a) and (b) indicates that the $k_y=0$ plane of the BZ can be accessed with a photon energy of $144$\,eV. (c) Shows the overlay of Fig.~$3$(a) and (b) to visualize the agreement of the DFT with the measurements. The white dashed line in (c) shows the corresponding cut location from (a) and (b)}
	\label{fig:SI_ARPES}
\end{figure}

For ARPES experiments, the crystal was cleaved in the $a-c$ plane along the natural platelet shape of the crystals.  With this cleavage plane, photon energy dependence probes the electronic structure along the $k_y$ axis ~\cite{Damascelli_2004}.  Fig.~\ref{fig:SI_ARPES}(a) shows a section of a photon energy-dependence sweep from $100$ to $184 eV$ in steps of $2 eV$, where spectra were integrated $\pm50 meV$ around the Fermi energy for each photon energy.  In Fig.~\ref{fig:SI_ARPES}(a), $k_{||}$ is along the BZ diagonal, in the $k_z$ --- $k_x$ plane, as indicated schematically in Fig.~\ref{fig:SI_ARPES} panel (a), (b), and (c) by a white dashed line. These are compared to calculated dispersions along the same cut, and yield qualitative agreement, though we note that photoemission matrix elements can be a function of photon energy and can cause features to be weaker or absent at some photon energies. Both data and theory have consistent features that largely do not disperse as a function of photon energy or $k_y$ for the chosen cut geometry, corresponding to FS$3$ and FS$4$. Near these minimally-dispersing features, but closer to the zone corner, is FS$5$, which moves closer to surfaces FS$3$ and FS$4$ at the $k_y = 0$ plane, and further away at the edge of the BZ. FS$2$, which is closer to the zone center, also moves closer to the minimally-dispersing surfaces at the $k_y = 0$ plane. This is one way we identify $144eV$ as the $k_y = 0$ plane. The other way we correspond photon energy with $k_y$ value is at the BZ boundaries. In the DFT calculation, there is more spectral weight at the BZ boundaries at $k_{||} = 0$, which is consistent with the enhanced intensity observed at 122 and 166 eV in ARPES data near $k_{||} = 0$.  In the main text, data are taken with a photon energy of 144 eV and cuts parallel to $k_x$.  In this experimental geometry, the most intense ARPES bands originate from FS$2$ and FS$3$. Fig.~\ref{fig:SI_ARPES}(c) shows an overlay of the data from the main text Fig. 3(a) and (b) in order to visualize the correspondence of the ARPES spectra to the DFT calculations in the $k_y = 0$ plane.

\section{Electronic structure}

\begin{figure}[!htb]
    \includegraphics[width=0.5\columnwidth]{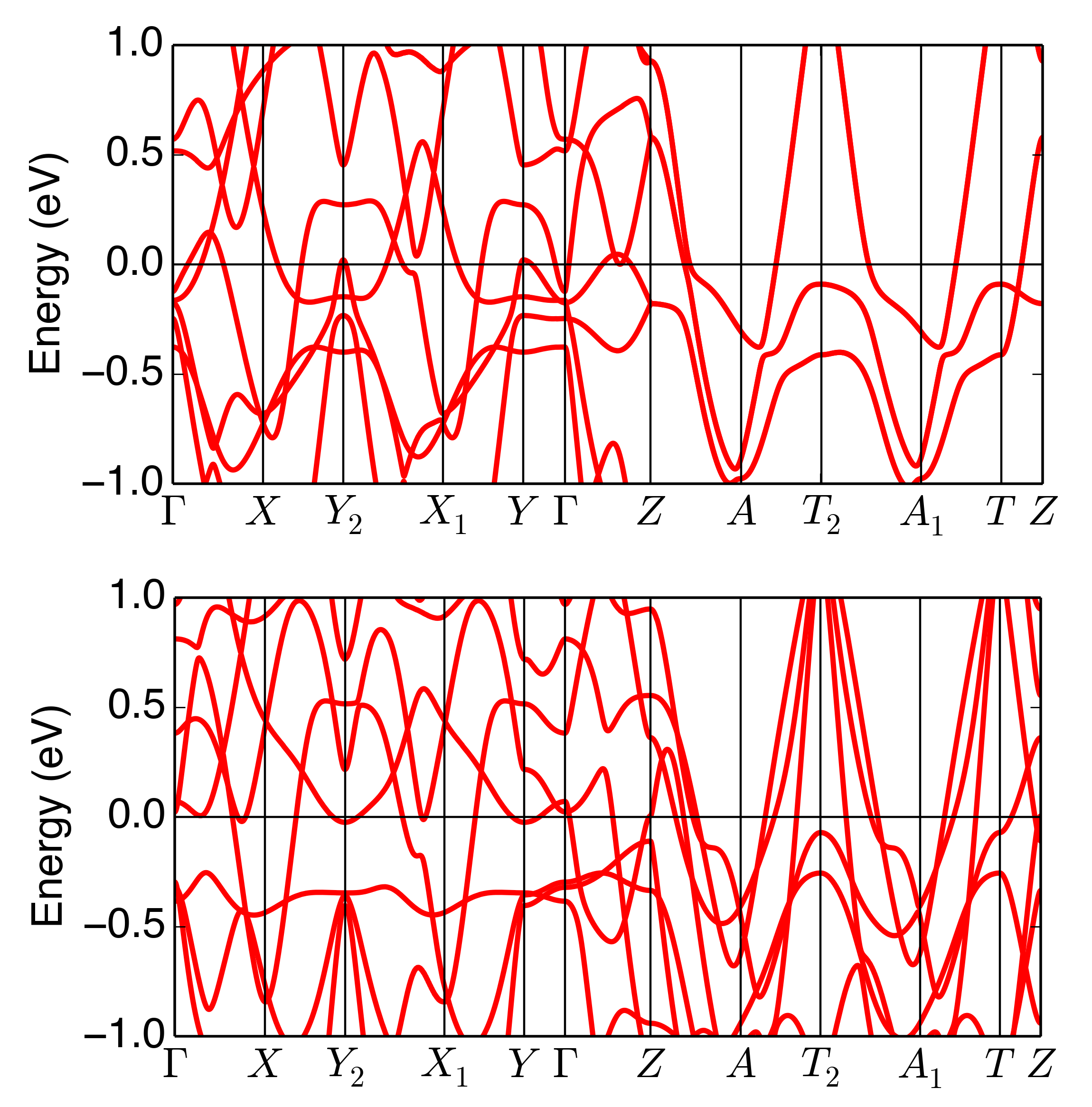}
    \caption{Band structure of nonsymorphic $Cmcm$ (top panel) versus symmorphic $Cmmm$ LaNiGa$_2$ (bottom panel), on a fine scale near E$_F$. For the symmetry point labels, see Fig.~2 of the main text. Fermi level band crossings are different for $Cmcm$, resulting in different Fermi surfaces than those shown by Singh \cite{Singh2012}.}
    \label{fig:bnd-compare}
\end{figure}

With the Ni $3d$ bands filled, no correlation correction beyond the generalized gradient approximation (GGA) is needed to describe the electronic structure. 
The FSs of LaNiGa$_2$ ($Cmcm$) are plotted in Fig.~$2$ of the main text to show evidence for the band degeneracies on the node surface, $k_z$=$\pi/c$ plane. The selected linear dispersions along $\vec{k}$=$(0,0.516\frac{\pi}{b},k_z)$ and $\vec{k}$=$(0.232\frac{\pi}{a},0,k_z)$ are also shown to provide evidence for the Dirac-like features of the loop and lines, respectively. To provide a clear picture of the effect of the nonsymmorphic symmetry operations on the band structure, we show both (without SOC) the band structure from the previous $Cmmm$ space group (bottom panel) and the updated $Cmcm$ unit-cell for LaNiGa$_2$ (top panel) in Fig.~\ref{fig:bnd-compare}.

As highlighted in the main text, both band structures have several bands which cross $E_F$ to produce five FSs. The key difference can be observed at and between the high-symmetry points on the $k_z$=$\pi/c$ plane. Here the bands, due to symmetry, `stick together' compared to the previous structure. As discussed above in the nonsymmorphic symmetry analysis section, this feature is directly the result of the previously undetected nonsymmorphic symmetry operations. 

A full report of the electronic structure around the nodal lines, and the effect of SOC, will be presented elsewhere.

\section{Comparison with other potential intrinsic topological superconductors}

Table~\ref{tab:TSCs} is a compiled list of potential intrinsic (single material) topological superconductors (TSCs). We indicate whether the topological features are reported for bulk bands and/or for surface states, if band crossings (BC) are located at $E_F$, if time-reversal symmetry breaking (TRS) is broken (B) or preserved (P) upon entering the superconducting state or if a magnetic (M) state preceded the superconductivity, if the space group is centrosymmetric (CS) or non-centrosymmetric (NC). As can be seen, LaNiGa$_2$ is unique in that it is the only intrinsic TSC which breaks time-reversal symmetry in the superconducting state and has topological features at $E_F$ without any overlapping magnetic state/fluctuations - antiferromagnetic (AFM) or ferromagnetic (FM). Additionally with our work, LaNiGa$_2$ is thus far the only material in this list to show experimental evidence for both broken time-reversal symmetry and topological features.

\begin{table*}[!htb]
	\centering
		\begin{tabular}{c c c c c cp{3.2cm}}
		\hline
		Material & Bulk or SSs? & at $E_F$? & TRS? & CS? & Refs. & Comments \\
		\hline
		\hline
			\rowcolor{Both}LaNiGa$_{2}$ & Bulk & $\checkmark$ & B & CS & \cite{Hillier2012} & \\
			\hline
			\rowcolor{Magnetic}UPt$_{3}$ & Both & $\checkmark$ & B & CS & \cite{Yanase2017,Schemm2014,Avers2020,Lee1993,Aeppli1988,Hayden1992,Luke1998} & AFM fluctuations\\
			\hline			
			\rowcolor{Magnetic}UCoGe, URhGe & Bulk & $\checkmark$ & M & CS & \cite{Daido2019,Fujimori2016,Aoki2019} & FM\\
			\hline
			\rowcolor{Magnetic}URu$_{2}$Si$_{2}$ & Bulk & $\times$ & M & CS & \cite{Oppeneer2010,Oppeneer2011,Das2012,Meng2013} & Hidden Order\\
			\hline
			\rowcolor{Magnetic}UTe$_{2}$ & - & - & B? & CS & \cite{Shishidou2021,Tokunaga2019,Sundar2019,Hayes2020} & FM-AFM fluctuations\\
			\hline
			\rowcolor{Cross}HfRuP & Bulk & $\checkmark$ & - & NC & \cite{Qian2019} & NLs cross $E_F$\\
			\hline
			\rowcolor{Cross}NbIr$_2$B$_2$,TaIr$_2$B$_2$ & Bulk & $\checkmark$ & - & NC & \cite{Gao2021} & NLs at $E_F$\\	
			\hline
			\rowcolor{Cross}NaAlSi & Bulk & $\checkmark$ & P & CS & \cite{Jin2019a,Muechler2019} & NLs cross $E_F$\\
			\hline
			\rowcolor{Cross}TaOsSi & Bulk & $\checkmark$ & - & CS & \cite{Xu2019} & NP at $E_F$\\
			\hline
			\rowcolor{Cross}MgB$_2$ & Bulk & $\checkmark$ & P & CS & \cite{Jin2019,Szabo2001} & NL crosses $E_F$\\
			\hline
			\rowcolor{Cross}CaSb$_2$ & Bulk & $\checkmark$ & - & CS & \cite{Funada2019,Ikeda2020,Takahashi2021} & NLs cross $E_F$ \\
			\hline
			\rowcolor{Cross}Zn$_5$Pt$_3$ & Bulk & $\checkmark$ & - & CS & \cite{Hamamoto2018,Bhattacharyya2021} & NLs cross $E_F$ \\
			\hline
			\rowcolor{Cross}SnTaS$_{2}$, PbTaS$_2$ & Both & $\checkmark$ & - & CS & \cite{Chen2019,Chen2021, Gao2020a} & NLs cross $E_F$\\
			\hline
			\rowcolor{Cross}Pb$_{1/3}$TaS$_2$ & Both & $\checkmark$ & - & CS & \cite{Yang2021} & NLs cross $E_F$\\
			\hline
			\rowcolor{Cross}In$_2$Bi & Both & $\checkmark$ & - & CS & \cite{Kuang2021} & NL cross $E_F$\\
			\hline
			\rowcolor{TRSB}Sr$_{2}$RuO$_{4}$ & - & - & B & CS & \cite{Luke1998,Veenstra2013,Sato2017,Pustogow2019Nature} & \\	
			\hline
			\rowcolor{TRSB}Sr$_{x}$Bi$_{2}$Se$_{3}$ & SS & $\times$ & B & CS & \cite{Han2015,Neha2019} & \\ 
			\hline
			Cu$_{x}$Bi$_{2}$Se$_{3}$, Nb$_{x}$Bi$_{2}$Se$_{3}$ & SS & $\times$ & - (P for Nb) & CS & \cite{Wray2010,Tanaka2012,Lahoud2013,Kobayashi2017,Das2020} & \\
			\hline
			Tl$_{x}$Bi$_{2}$Te$_{3}$ & SS & $\times$ & - & NC & \cite{Trang2016} & \\
			\hline
			$\alpha$-PdBi$_{2}$ & SS & $\times$ & - & CS & \cite{Dimitri2018} & \\
			\hline
			$\alpha$-BiPd & SS & $\times$ & - & NC & \cite{Sun2015,Neupane2016,Pramanik2020} & \\
			\hline
			$\beta$-PdBi$_{2}$ & SS & $\times$ & P & CS & \cite{Sakano2015,Biswas2016} & \\
			\hline
			FeTe$_{1-x}$Se$_{x}$ & SS & $\times$ & P$^{\dagger}$ & CS & \cite{Zhang2018a,Zhang2019b,Biswas2010} & \\
			\hline
			LiFeAs & Both & $\times$ & P & NC & \cite{Zhang2019b,Wright2013} & \\
			\hline
			(Li$_{0.84}$Fe$_{0.16}$)OHFeSe & SS & $\times$ & - & NC & \cite{Liu2018} & \\
			\hline
			CaKFe$_{4}$As$_{4}$ & SS & $\times$ & - & CS & \cite{Liu2020} & \\
			\hline
			In$_{x}$TaS$_{2}$, In$_{x}$TaSe$_{2}$ & Bulk & $\times$ & - & NC & \cite{Li2020a,Li2021} & \\
			\hline
			Cu$_{x}$ZrTe$_{1.2}$ & Bulk & $\times$ & - & CS & \cite{Machado2017} & \\
			\hline
			NiTe$_{2}$, PdTe$_2$ & Both & $\times$ & - (P for Pd) & CS & \cite{Huang2016,Fei2017,Noh2017,Singh2019a,Zhang2020b} & \\
			\hline
			PbTaSe$_{2}$ & Both & $\times$ & P & NC & \cite{Chang2016,Chen2016b,Bian2016,Guan2016,Wilson2017} & \\
			\hline
			NbC, TaC & Bulk & $\times$ & P & CS & \cite{Shang2020b,Yan2020} & \\
			\hline
			Mo$_{2}$C, W$_2$C & SS & $\times$ & - & CS & \cite{Zhao2020} & \\
			\hline
			CaSn$_3$, BaSn$_3$ & Bulk & $\times$ & - & CS & \cite{Gupta2017,Zhang2020a} & \\
			\hline
			YRuB$_{2}$, LuRuB$_2$ & Bulk & $\times$ & P & CS & \cite{Barker2018,Gao2020} & \\
			\hline
			YIn$_{3}$ (M=In,Pb,Tl) & SS (Both for Tl) & $\times$ & -  & CS & \cite{Tu2019} & \\
			\hline
			NbAl$_{3}$ & Bulk & $\times$ & - & CS & \cite{Chen2018a} & \\
			\hline
			TaSe$_{3}$ & SS & $\times$ & - & CS & \cite{Nie2018,Xia2020} & \\
			\hline
			Ta$_{3}$Sb, Ta$_3$Sn & SS (Both for Sn) & $\times$ & - & CS & \cite{Bradlyn2016,Kim2019,Derunova2019} & \\
			\hline
			Nb$_{3}$M (M=Al,Os,Au) & Bulk & $\times$ & - & CS & \cite{SreenivasaReddy2016,Naher2018,Derunova2019} & \\
			\hline
			LaNiSi, LaPtSi, LaNiGe & Bulk & $\times$ & P & NC & \cite{Zhang2020,Sajilesh2018,Sajilesh2020} & \\
			\hline
			TlBiTe$_{2}$ & SS & $\times$ & - & CS & \cite{Chen2010a} & \\
			\hline
			Tl$_{5}$Te$_{3}$ & SS & $\times$ & - & CS & \cite{Arpino2014} & \\
			\hline
			YPtBi, LuPtBi & SS & $\times$ & - (P for Y) & NC & \cite{Liu2016,Hosen2020,Bay2014,AlSawai2010,Nakajima2015} & other half-Heuslers\\
			\hline
			KV$_{3}$Sb$_{5}$, CsV$_{3}$Sb$_{5}$ & Both & $\times$ & - (P for Cs) & CS & \cite{Ortiz2019,Ortiz2020,Ortiz2021,Gupta2021} & \\
			\hline
			Sn$_{1-x}$In$_{x}$Te & SS & $\times$ & P & CS & \cite{Sato2013,Saghir2014,Polley2016,Schmidt2020,Schmidt2020a} & \\
			\hline
			MoTe$_2$ & Bulk & $\times$ & - & NC & \cite{Sun2015a,Deng2016,Huang2016a,Jiang2017} & $T_{d}$ phase\\
			\hline
			WS$_{2}$ & SS & $\times$ & - & CS & \cite{Fang2019} & $2$M phase.\\
			\hline
			A$_2$Cr$_3$As$_3$ (A=Na,K,Rb,Cs) & Bulk & $\times$ & - (P* for K) & NC & \cite{Xu2020,Adroja2015} & \\
			\hline
			ZrInPd$_2$, HfInPd$_2$ & Bulk & $\times$ & - & CS & \cite{Mondal2019} & \\
			\hline
			MM'$_2$Al (M=Zr,Hf; M'=Ni,Pd)& Bulk & $\times$ & - & CS & \cite{Guo2017a} & \\
			\hline
			MPd$_2$Sn (M=Sc,Y,Lu) & Both & $\times$ & - (P for Y) & CS & \cite{Saadaoui2013,Guo2017a,Guo2018} & \\
			\hline
		\end{tabular}
	\caption{Compiled list of potential intrinsic (single material) TSC materials. All properties are exhibited at ambient pressure. NP: nodal point, NL: nodal line. The highlighted red rows are materials with band-crossings at $E_F$, the orange rows show the U-based materials which have overlapping magnetic ordering/fluctuations and superconductivity, the blue rows show materials which breaks time-reversal symmetry, and the purple row shows LaNiGa$_2$ as the only material to date with broken time-reversal symmetry and band-crossings at $E_F$. $^{\dagger}$FeTe$_{1-x}$Se$_x$: recently reported to break time-reversal symmetry using a method other than $\mu$SR or Polar Kerr effect~\cite{Zaki2021}. *K$_2$Cr$_3$As$_3$: possible very weak internal field (0.003\,G)~\cite{Adroja2015}.}
	\label{tab:TSCs}
\end{table*}

\newpage
\bibliographystyle{apsrev4-2}
\bibliography{Library,biblio}